\documentstyle[twoside,fleqn,espcrc2,epsfig]{article}
\newcommand{\gsim}{\raisebox{-0.07cm}{$\,\stackrel{>}{{\scriptstyle
 \sim}}\, $} }
\newcommand{\lsim}{\raisebox{-0.07cm}{$\,\stackrel{<}{{\scriptstyle
 \sim}}\, $} }
\newcommand\qV{\mbox{\boldmath $q$}}
\newcommand\Fvec{\,\mbox{\boldmath $F$}}
\newcommand\Gvec{\,\mbox{\boldmath $G$}}
\newcommand\GGV{\,\mbox{\boldmath $\gamma$}}

\newcommand\PV{\mbox{\boldmath $P$}}
\newcommand\MV{\,\mbox{\boldmath $M$}}

\newcommand\UV{\,\mbox{\boldmath $U$}}
\newcommand\GA{\,\mbox{\boldmath $\Gamma$}}
\newcommand\gV{\,\mbox{\boldmath $\gamma$}}

\newcommand\MeV{\,\mbox{MeV}}

\newcommand\beq{\begin{equation}}
\newcommand\eeq{\end{equation}}
\newcommand\bea{\begin{eqnarray}}
\newcommand\eea{\end{eqnarray}}
\newcommand\MSbar{$\overline{\mbox{MS}}$}
\newcommand\ra{\rightarrow}

\title{$\!\! $The Evolution of Unpolarized and Polarized
 Structure Functions at Small $x${\thanks{Talk presented by A. Vogt
 }}$\!\!\!\! $}
\author
{J. Bl\"umlein$^{\,\rm a}$, S. Riemersma
 \address{DESY--Zeuthen, Platanenallee 6, D--15735 Zeuthen, Germany},
and A. Vogt
 \address{Institut f\"ur Theoretische Physik, Universit\"at W\"urzburg,
          Am Hubland, D--97074 W\"urzburg, Germany}
}

\begin{document}
\begin{titlepage}

\large
\begin{flushleft}
DESY 96--131 \\[0.1cm] WUE--ITP--96--016 \\[0.1cm] August 1996
\end{flushleft}
\vspace{0.4cm}
\begin{center}
\LARGE
{\bf The Evolution of Unpolarized and Polarized} \\
\vspace{0.1cm}
{\bf Structure Functions at Small ${\boldmath x}^{\ast} $} \\
\vspace{1.4cm}
\large
J. Bl\"umlein, S. Riemersma \\
\vspace{0.5cm}
\large {\it
DESY--Zeuthen \\
\vspace{0.1cm}
Platanenallee 6, D--15735 Zeuthen, Germany }\\
\vspace{1.0cm}
\large
A. Vogt\\
\vspace{0.5cm}
\large {\it
Institut f\"ur Theoretische Physik, Universit\"at W\"urzburg \\
\vspace{0.1cm}
Am Hubland, D--97074 W\"urzburg, Germany} \\
\vspace{1.6cm}
{\bf Abstract}
\end{center}
\vspace{-0.1cm}
A survey is given of recent developments on the resummed small-$x$ 
evolution, in a framework based on the renormalization group equation, 
of non--singlet and singlet structure functions in both unpolarized
and polarized deep--inelastic scattering. The available resummed
anomalous dimensions are discussed for all these cases, and the most
important analytic and numerical results are compiled. The quantitative
effects of these small-$x$ resummations on the evolution of the various
parton densities and structure functions are presented, and their
present uncertainties are investigated. An application to QED radiative
corrections is given.
\vfill 
\noindent
\normalsize
$^{\ast} $ Based on invited talks presented by J. Bl\"umlein at the 
{\sf International Workshop on Deep Inelastic Scattering and Related
Phenomena (DIS'96)\/}, Rome, Italy, April~1996; and by A. Vogt at the
{\sf 1996 Zeuthen Workshop on Elementary Particle Theory: QCD and QED
in Higher Orders\/}, Rheinsberg, Germany, April~1996. To appear in the 
Proceedings of the
Rheinsberg Workshop, Nucl.\ Phys.\ {\bf B} (Proc.\ Suppl.).
 
\end{titlepage}
%
%
\begin{abstract}
\noindent
A survey is given of recent developments on the resummed small-$x$ 
evolution, in a framework based on the renormalization group equation, 
of non--singlet and singlet structure functions in both unpolarized
and polarized deep--inelastic scattering. The available resummed
anomalous dimensions are discussed for all these cases, and the most
important analytic and numerical results are compiled. The quantitative
effects of these small-$x$ resummations on the evolution of the various
parton densities and structure functions are presented, and their
present uncertainties are investigated. An application to QED radiative
corrections is given.
\vspace*{-1.4mm}
\end{abstract}
\maketitle

\section{Introduction}

\vspace*{-1mm}
\noindent
The evolution kernels of both non--singlet and singlet, unpolarized and
polarized parton densities contain large logarithmic contributions for
small fractional momenta $x$. For unpolarized deep--inelastic scattering
(DIS) processes the leading small-$x$ contributions in the singlet case
behave like
\cite{LIPAT}
$$
 \left ( \frac{\alpha_s}{N - 1} \right )^k
 \:\:\: \leftrightarrow \:\:\:
 \frac{1}{x} \,  \alpha_{s}^{k} \, \ln^{k-1} x \:\: ,
\vspace*{-0.5mm}
$$
where $N$ is the Mellin variable. The leading terms for the unpolarized
and polarized non--singlet and the polarized singlet cases are
of the form~\cite{KL,BER2}
$$
 N \left (  \frac{\alpha_s}{N^2} \right )^k
 \:\:\: \leftrightarrow \:\:\:
 \alpha_{s}^{k} \, \ln^{2k-2} x \:\: .
\vspace*{-0.5mm}
$$

The resummation of these terms to all orders in the strong coupling
constant $\alpha_s$ can be completely derived by means of perturbative
QCD. Since infinities, such as the ultraviolet and collinear
divergencies, emerging in the calculation of the higher--order
corrections have to be dealt with, the {\it only} appropriate
framework for carrying out these resummations is provided by the
renormalization group equations.
The impact of the resulting all--order anomalous dimensions on the
behaviour of the DIS structure functions at small $x$  depends as
well on the non--perturbative input parton densities at an initial
scale $Q_0^2$. Thus the resummation effects can only be studied via
the evolution over some range in $Q^2$.

This evolution moreover probes the anomalous dimensions also at medium
and large values of $x$ by the Mellin convolution with the parton
densities. Hence the small-$x$ dominance of the leading terms over
contributions less singular as $x \ra 0 $ in the anomalous dimensions
does {\it not\/} necessarily imply the same situation for observable
quantities, such as the structure functions.
These aspects need to be considered to arrive at sound conclusions
about the consequences of the small-$x$ resummations on physical
quantities.

In the present paper we give a survey of the recent developments in
this field. The general framework for the evolution of parton densities
and structure functions is recalled in Section 2.
Section 3 reviews the available results on the resummed anomalous
dimensions for the various DIS processes. Numerical coefficients for
their expansions in $\alpha _s$ are compiled, as well as the analytical
predictions for the most singular contributions to the 3--loop splitting
functions. The issue of subleading terms is discussed, guided by the 
known 2--loop results.
In Section~4 the numerical implication of these resummations are
investigated for the various unpolarized and polarized cases. The
uncertainties due to possible less singular terms and due to
insufficiently constrained initial parton densities are illustrated.
An application to QED radiative corrections is presented.
Section~5 summarizes the main results.

\section{The evolution equations}
\label{sect2}

\noindent
The twist-2 contributions to any deep--inelastic scattering structure
function can be represented in general, see e.g.~\cite{FP,CH},
by the three flavour non--singlet combinations of quark densities
\bea
\label{NSpar1}
 q_{{\rm NS},i}^{\pm} &\! =\! &  q_i \pm \overline{q}_i - \frac{1}{N_f}
 \sum_{r=1}^{N_f} (q_r \pm \overline{q}_r) \: , \\
 \label{NSpar2}
 q_{\rm NS}^{\rm val} &\! =\! & \sum_{r=1}^{N_f} (q_r - \overline{q}_r)
 \: ,
\eea
and the  singlet quark and gluon
distributions
\beq
\label{sing}
 \qV_S= \left( \begin{array}{c} \!\Sigma\! \\ \! g\! \end{array}\right)
 \:\: ,\:\: \Sigma \equiv \sum_{r=1}^{N_f} \, (q_{r}+\bar{q}_{r}) \: .
\eeq
Here $N_f$ denotes the number of active (massless) quark flavours.
The structure functions $F_i(x,Q^2)$ are obtained by
\bea
\label{strf}
 F_i(x,Q^2) &\! =\! &
 \sum_{r=1}^{2N_f} a_{ir} \, c_{i,r}(x,Q^2) \otimes q_r(x,Q^2)
 \nonumber \\
 & & \mbox{} + a_{ig} \, c_{i,g} (x,Q^2) \otimes g(x,Q^2) \: ,
\eea
where the factors $a_{ij}$ depend on the electroweak couplings, and
$c_{i,j}(x,Q^2)$ denote the respective coefficient functions. Finally
$\otimes $ stands for the Mellin convolution in the first variable,
\bea
\label{Mellin}
 \lefteqn{A(x) \otimes B(x) = } \\
 & & \int_0^1 \! dx_1 \int_0^1 \! dx_2 \, \delta(x - x_1 x_2)
 \, A(x_1) B(x_2) \nonumber \: .
\eea

The above notation is used in a  generic way for both unpolarized and
polarized DIS, i.e.\ for the polarized case the replacements
\beq
\label{REPL}
 q \rightarrow \Delta q ~,~~~\overline{q} \rightarrow \Delta
 \overline{q} ~,~~~{\rm and}~~ g \rightarrow \Delta g,
\eeq
are understood with, e.g., $ \Delta q $ given in terms of the spin
projections $q\! \uparrow $ and $q\! \downarrow $ via
\beq
\label{DEL }
 \Delta q = q\!\uparrow -  q\!\downarrow \: .
\eeq
Correspondingly the splitting functions (see below) and the coefficient
functions in eq.~(\ref{strf}) have to be replaced.

As long as the splitting functions $P_{qq}^S$ and $P_{q\overline{q}}^S$
(cf.\ ref.~\cite{FP}) do not differ, the evolution equations are
identical for the combinations $q_{{\rm NS},i}^{-}$ and $q_{{\rm NS}}
^{\rm val}$. Since this is the case at all orders known presently
(i.e.\ up to next-to-leading order, NLO)\footnote{So far only for the
anomalous dimension $\gamma^+_{\rm NS}$ the first moments have been
calculated to 3--loop order~\cite{LOOP3}.}, we will not investigate
the evolution of the combination (\ref{NSpar2}) separately in the
following.

The evolution equations for the non--singlet and singlet combinations
of the parton distributions are then given by
\bea
\label{evol1}
 \frac{\partial\, q_{\rm NS}^{\,\pm}(x,Q^2)}{\partial \ln  Q^2} &=&
  P_{\rm NS}^{\,\pm}(x, \alpha_s ) \otimes q_{\rm NS}^{\,\pm}(x,Q^2)
 \: , \nonumber \\
\label{evol2}
 \frac{\partial\, \qV _{\rm S}(x,Q^2)}{\partial \ln  Q^2} &=&
  \PV _{\rm S} (x, \alpha_s ) \otimes \qV _{\rm S}(x,Q^2) \: .
\eea
The splitting functions $P_{\rm NS}^{\,\pm}$ and $\PV _{\rm S} $ are
specified below. Note that the unpolarized (polarized) quark density
combinations $q_{\rm NS}^-$ and $q_{\rm NS}^+$ evolve with $P_{\rm NS}
^{-}$ ($P_{\rm NS}^{+}$) and $P_{\rm NS}^{+}$ ($P_{\rm NS} ^{-}$),
respectively.

In the following, we will simplify the notation by dropping the
subscripts `NS' and `S', and use the abbreviation $a_s \equiv
\alpha_s(Q^2) /4 \pi $ for the running QCD coupling for convenience.
The scale dependence of $a_s$ is governed by
\beq
 \label{eval}
 \frac{da_s}{d\ln Q^2} = - \sum_{k=0}^{\infty} a^{k+2}_s \beta_k \: ,
\eeq
where only $ \beta_0 = (11/3)\, C_A - (4/3)\, T_F N_f $ and
$ \beta_1 = (34/3)\, C_A^2 - (20/3)\, C_A T_F N_f -  4\, C_F T_F N_f$
enter up to NLO. The colour factors are $ C_F = (N_c^2-1)/(2 N_c)
\equiv 4/3 $, $ C_A = N_c \equiv 3 $, $ T_F = 1/2 $.
For the numerical calculations in Section~4 we use
\beq
 \label{als}
a_s = \frac{1}{\beta_0} \ln(Q^2/\Lambda^2) \left [ 1
- \frac{\beta_1 \ln \ln(Q^2/\Lambda^2)}{\beta_0^2 \ln(Q^2/\Lambda^2)}
\right ],
\eeq
with $\Lambda$ being the QCD scale parameter.
The splitting functions and  coefficient functions can be represented
by the series
\bea
\label{Pns}
 P^{\pm}(x,a_s)
   &\! =\! &  \sum_{l=0}^{\infty} a_{s}^{l+1} P_l^{\pm}(x) 
   \: , \nonumber \\
\label{Ps}
 \PV (x,a_s)
   &\!\equiv \!& \left( \begin{array}{cc}
               \! P_{qq}(x,a_s) \! & \! P_{qg}(x,a_s) \! \\
               \! P_{gq}(x,a_s) \! & \! P_{gg}(x,a_s) \!
                        \end{array} \right)  \\
   &\! =\! &   \sum_{l=0}^{\infty} a_s^{l+1} \PV_{l}(x) 
   \: , \nonumber \\
\label{coef}
 c_{i,j}(x,Q^2) &\! =\! & \delta(1-x)\delta_{jq}
		        + \sum_{l=1}^{\infty} a_s^l c_{ij,l}(x) \: .
\eea
Unless another scheme is stated explicitly, we will always refer to the
\MSbar\ scheme both for renormalization and factorization, and take
$Q^2$ as the renormalization and factorization scale.

$\!\!$The expansion coefficients $ P^{-}_{l}(x)$ and $\PV_{l}^{\rm
unpol} (x)$ are subject to the sum rules
\bea
\label{conserv}
  & & \int_0^1 \! dx \, P_{l}^-(x) = 0 \: , \nonumber \\
  & & \int_0^1 \! dx \, x \sum_i P_{ij,l}^{\rm unpol.}(x) = 0 \: ,
\eea
which are due to fermion number and energy momentum conservation,
respectively. By now all the unpolarized and polarized splitting
functions are completely known up to NLO, $l=1$. The full expressions
for their $x$--dependences can be found in
refs.~\cite{SpF1}--\cite{SpF5}. The most singular contributions as
$x \rightarrow 0$ will be displayed in Section~\ref{sect34}.

The parton densities are not observables beyond leading order. Hence
it is convenient to consider also the evolution equations for related
physical quantities, the structure functions $F_i(x,Q^2)$, directly.
In non--singlet cases, the all--order resummation of the leading
small-$x$ contributions has in fact been given \cite{KL} on the level
of the structure function combinations $F^{\pm}_i(x,Q^2)$.
Their evolution equations read
\beq
\label{evolf}
 \frac{\partial \, F_i^{\pm}(x, a_s)}{\partial a_s} = - \frac{1}
 {\beta_0 a_s^2} K_i^{\pm}(x, a_s) \otimes F_i^{\pm}(x, a_s) \,
\eeq
after a transformation to an equation in $a_s$. Here, e.g., the NLO
kernels can be written as
\bea
\label{KNS}
 \lefteqn{K_{i,1}^{\pm}(x, a_s) = } \\
  & & \hspace*{-3mm} a_s \, P_{0}(x) + a_s^2 \left[ P_1^{\pm}(x)-
  \frac{\beta_1}{\beta_0} P_{0}(x) - \beta_0 c_{i,1}^{\pm}(x) \right] ,
  \nonumber
\eea
with $c_{i}^{\pm}(x)$ denoting the corresponding coefficient function
combinations.
Generally the terms $\propto a_s (a_s \ln^2 x)^l $ emerge in the
$a_s$ expansion of the kernels $ K_{i}^{\pm}(x, a_s) $ only in
combination with the coefficient $\beta_0$. As will be outlined below
the leading small-$x$ contributions to these kernels coincide with
those of the \MSbar\ splitting functions $P_l^{\pm}(x)$, since the
corresponding coefficient functions $c_{i,l}^{\pm}(x)$ turn out to be
less singular as $x \rightarrow 0$.

\section{Resummation of leading small-$x$ terms}
\label{sect3}
\subsection{The non-singlet case}
\label{sect31}

\vspace*{1mm}
\noindent
The most singular contributions to the Mellin transforms of the
all--order evolution kernels $K^{\pm}(x,a_s)$ for the non--singlet
structure functions have been obtained via
\bea
\label{Kpm}
 \lefteqn{ {\cal M}\left[ K^{\pm}_{x \rightarrow 0}(a_s) \right ] (N)
 \equiv  \int_0^1 \! dx \, x^{N-1} K_{x \rightarrow 0}^{\pm}(x, a_s)}
 \nonumber \\
 & & \equiv  - \frac{1}{2} \GA_{x \rightarrow 0}^{\pm}(N, a_s) =
 \frac{1}{8\pi ^2} f_0^{\pm}(N, a_s)
\eea
from  positive and negative signature amplitudes $f_0^{\pm}(N, a_s)$
studied in ref.~\cite {KL} (cf.\ Section 3.2):
\[
  \GA_{x \rightarrow 0}^+(N,a_s) = -N
  \left \{ 1 - \sqrt{1 - \frac{8 a_s C_F}{N^2}} \right \} \: , 
\]
\vspace{-5mm}
\bea
\label{Gns}
 \lefteqn{ \GA_{x \rightarrow 0}^-(N,a_s) = } \\
  & & \hspace*{-3mm} -N \left \{ 1 - \sqrt{1 - \frac{8 a_s C_F}{N^2}
  \left [1 - \frac{f_8^{+}(N,a_s)}{2\pi^2 N} \right ] } \right \} \: .
  \nonumber
\eea
Here the colour octet amplitude $f_8^{+}(N,a_s)$ reads
\bea
\label{f8}
 \lefteqn{ f_8^{+}(N,a_s) = 16 \pi^2 N_c a_s \frac{d}{d N}
  \ln \left [ e^{\, z^2/4} D_{-1/[2N_c^2]}(z) \right ] } \nonumber \\
 & &
\eea
with $z = N/ \sqrt{2 N_c a_s}$ and $D_p(z)$ denoting the parabolic
cylinder function \cite{RYGRA}.

\begin{table}[b]
\vspace{-10mm}
\begin{center}
\small
\begin{tabular}{||r|l|l||}
\hline \hline
 & & \\[-0.3cm]
\multicolumn{1}{||c|}{$l$} &
\multicolumn{1}{c|} {$K_l^{+}$} &
\multicolumn{1}{c||}{$K_l^{-}$} \\
 & & \\[-0.3cm] \hline \hline
 & & \\[-0.3cm]
      0 &  ~2.667E0   & ~2.667E0  \\
      1 &  ~3.556E0   & ~5.333E0  \\
      2 &  ~1.580E0   & ~1.432E0  \\
      3 &  ~3.512E-1  & ~9.964E-1 \\
      4 &  ~4.682E-2  & -2.078E-1 \\
      5 &  ~4.162E-3  & ~1.448E-1 \\
      6 &  ~2.643E-4  & -5.777E-2 \\
      7 &  ~1.258E-5  & ~2.168E-2 \\
      8 &  ~4.661E-7  & -7.173E-3 \\
      9 &  ~1.381E-8  & ~2.143E-3 \\
     10 &  ~3.348E-10 & -5.827E-4 \\
\hline \hline
\end{tabular}
\normalsize
\end{center}

\noindent
{\sf {\bf Table~1:}~~The coefficients $K^{\pm}_l$ of the expansion of
 $ K^{\pm}_{x \rightarrow 0}(x,a_s)$ in terms of $a_s (a_s \ln^2 x)^l$
 as obtained from the resummations in eqs.\ (\ref{Gns}) and
 (\ref{f8}).}
\vspace{-1mm}
\end{table}

Expanding the resummed anomalous dimensions (\ref{Gns}) into a series
in $a_s$ and transforming back to $x$--space yields the numerical
coefficients shown in Table 1.
The first two terms of the resummed kernel
$K^{\pm}_{x \rightarrow 0}(x, a_s)$
agree with the leading small-$x$ contributions of the corresponding
LO and NLO splitting functions $P^{\pm} (x, a_s)$ \cite{BVplb1}.
This expansion also reveals a significant theoretical difference between
$\GA_{x \rightarrow 0}^+(N,a_s)$ and $\GA_{x \rightarrow 0}^-(N,a_s)$:
the $a_s$ series is convergent in the former but not in the latter case,
where it involves the asymptotic expansion of $D_p(z)$.

Unlike the splitting functions, the coefficient functions $c_i^{\pm}
(x,a_s)$ are known up to next-to-next-to-leading order (NNLO), $l=2$,
in the \MSbar\ scheme \cite{CO1,CO2}. At small $x$ they rise only as
\beq
\label{coef2}
 c^{\pm}_{i,1} \propto  \ln x \: , \:\:\:
 c^{\pm}_{i,2} \propto  \ln^3 x \: .
\eeq
Therefore the third expansion coefficient of $K^{\pm}_{x \rightarrow 0}
(x, a_s)\, $ leads to a prediction for the most singular parts of the
non--singlet \MSbar\ splitting functions in NNLO, $P_{2}^{\pm}(x)$,
given by \cite{BVplb1}
\beq
  P_{2,\, x \rightarrow 0}^{+}(x) = \frac{2}{3} C_F^3 \ln^4 x \: , 
\eeq
$$
 P_{2,\, x \rightarrow 0}^{-}(x) =
 \left[ -\frac{10}{3} C_F^3 + 4 C_F^2 C_A - C_F C_A^2 \right]
 \ln^4 x \: .
$$
All the methods of this section have been applied to QED and also
comparisons with the available fixed order calculations were carried
out. For details we refer to refs.~\cite{BVcrac,BRVqed}.

\vspace*{1mm}
\subsection{The polarized singlet case}
\label{sect32}

\vspace*{1mm}
\noindent
The amplitude relations  of ref.~\cite{KL} have been generalized
to the polarized singlet case recently \cite{BER2}:
\bea
\label{eq31}
\lefteqn{ \Fvec_0(N,a_s) = 16 \pi^2 \frac{a_s}{N} \MV_0 } \\
 & & - \frac{8 a_s}{N^2} \Fvec_8(N,a_s) \Gvec_0 + \frac{1}{8 \pi^2}
 \frac{1}{N} \Fvec_0^2(N,a_s) \: , \nonumber \\
\lefteqn{ \Fvec_8(N,a_s) = 16 \pi^2 \frac{a_s}{N} \MV_8 } \\
  & & + \frac{2 a_s}{N} C_G \frac{d}{d N} \Fvec_8(N,a_s)
  + \frac{1}{8 \pi^2} \frac{1}{N} \Fvec_8^2(N,a_s) \: . \nonumber
\label{eq32}
\eea
Here the basic  matrices are given by
\[
 \MV_0 = \left( \begin{array}{cc} \! C_F \!\! & -2 T_F N_f\! \\
                \! 2 C_F \!\! &  4 C_A  \!  \end{array} \right),
 \Gvec_0 = \left ( \begin{array}{cc} \! C_F \! & 0   \! \\
             \! 0 \! & C_A \! \end{array} \right) ,
\]
\vspace{-3mm}
\beq
\label{M0M8}
 \MV_8 = \left ( \begin{array}{cc} \! C_F - C_A/2 &  -T_F N_f \! \\
                 \!  C_A &  2 C_A \!  \end{array} \right).
\eeq
Note that the matrix $\MV_0$ is the $x \rightarrow 0$ limit of the
well--known matrix of the polarized splitting functions in
LO~\cite{SpF2}.
The equations for $f_0^{\pm}(N,a_s) $ and $ f_8^{+}(N,a_s) $ of Section
3.1 are entailed in these expressions by keeping only the $qq$-entries
of the matrices (\ref{M0M8}), and, in the `+'-case, additionally
dropping the $\Fvec_8$-term in eq.~(\ref{eq31}). Also for the polarized
singlet structure function $g_1 (x,Q^2)$ the coefficient functions are
known up to NNLO~\cite{CO2}, and their leading small-$x$ behaviour is
the same as in eq.~(\ref {coef2}). The resummed leading small-$x$
contributions to the splitting functions are thus related to
$\Fvec_0 (N,a_s)$ by
\bea
\label{eqSER}
 \PV(x, a_s)_{x \rightarrow 0}
  &\! = \! &  \sum_{l=0}^{\infty} \PV^{\, x \rightarrow 0}_{l}(x)\,
  a_s^{l+1}  \\
  &\! = \! & \frac{1}{8\pi^2} {\cal M}^{-1}\left[ \Fvec_0(N, a_s)
  \right] (x) \: . \nonumber
\eea

Eqs.~(\ref{eq31}) and (\ref{eq32}) have been solved directly in terms of
a series in $a_s$ in ref.\ \cite{BVplb2}. Unlike the previous non--%
singlet case, the representation (\ref{eqSER}) is needed for the
analytical $N$-space solution of the evolution equations here, cf.\
Section~4.3. Also in the present case the lowest--order expansion
coefficients $\PV^{\, x \rightarrow 0}_{0,1}(x)$ agree with the
corresponding limit of the NLO splitting function matrices \cite{BER2}.
The predictions for the NNLO quantities $P_{ij,2}^{\, x \rightarrow 0}$
read \cite{BVplb2}:
\bea
 P_{qq,2}^{\, x \rightarrow 0}(x) &\!\!\!\! =\!\! & \!\!\!
 \frac{2}{3} C_F \Big[
  -5\, C_F^{2}-\frac{3}{2}\, C_A^{2}+6\,{\it C_A}\, C_F \nonumber \\
  & &\mbox{} -8\,{\it T_F N_f }\, C_F-6{\it T_F N_f}\,{\it C_A} \Big]
  \ln^4 x \: , \nonumber\\
 P_{qg,2}^{\, x \rightarrow 0}(x) &\!\!\!\! =\!\! & \!\!\!
 \frac{2}{3} T_F N_f\Big[
  -15\, C_A^{2}+2\, C_F^{2}-6\,{\it C_F}\,{\it C_A} \nonumber \\
  & &\mbox{} +8\,T_F N_f C_F \Big] \ln^4 x
 \: , \\
 P_{gq,2}^{\, x \rightarrow 0}(x) &\!\!\!\! =\!\! & \!\!\!
 \frac{2}{3} C_F \Big[
  15\, C_A^{2}-2\, C_F^{2}+6\,{\it C_F}\,{\it C_A} \nonumber \\
  & &\mbox{} -8\,T_F N_f C_F \Big] \ln^4 x
 \: , \nonumber\\
 P_{gg,2}^{\, x \rightarrow 0}(x) &\!\!\!\! =\!\! & \!\!\!
 \frac{2}{3} \Big[
  28\, C_A^{3}+2\, T_F N_f\, C_A^{2}- 4 T_F N_f\, C_F^{2} \nonumber\\
  & &\mbox{} -24\, C_F\, T_F N_f\, C_A  \Big] \ln^4 x \, . \nonumber
 \label{eqP2}
\eea
The expansion coefficients up to $l=10$ are shown in a compact
numerical form in Table 2. For a more precise  representation see
ref.~\cite{BVplb2}.

\begin{table}[t]
\vspace*{-2mm}
\begin{center}
\small
\begin{tabular}{||r|l|l|l|l||}
\hline \hline
\multicolumn{1}{||c|}{ } &
\multicolumn{4}{  c||}{ } \\[-0.3cm]
\multicolumn{1}{||c|}{   } &
\multicolumn{4}{  c||}{$N_f = 3$   } \\
\multicolumn{1}{||c|}{ } &
\multicolumn{4}{  c||}{ } \\[-0.3cm] \hline
 & & & & \\[-0.3cm]
\multicolumn{1}{||c|}{$l$} &
\multicolumn{1}{c|}{$P_{qq}^{(l)}$} &
\multicolumn{1}{c|}{$P_{qg}^{(l)}$} &
\multicolumn{1}{c|}{$P_{gq}^{(l)}$} &
\multicolumn{1}{c||}{$P_{gg}^{(l)}$} \\
 & & & & \\[-0.3cm] \hline \hline
 & & & & \\[-0.3cm]
  0 &  ~2.667E0 & -6.000E0 & ~5.333E0 &  2.400E1 \\
  1 &  -1.067E1 & -4.400E1 & ~3.911E1 &  1.280E2 \\
  2 &  -3.679E1 & -1.394E2 & ~1.240E2 &  4.189E2 \\
  3 &  -6.642E1 & -2.288E2 & ~2.034E2 &  6.981E2 \\
  4 &  -6.110E1 & -2.154E2 & ~1.915E2 &  6.685E2 \\
  5 &  -3.858E1 & -1.347E2 & ~1.197E2 &  4.201E2 \\
  6 &  -1.680E1 & -5.955E1 & ~5.294E1 &  1.868E2 \\
  7 &  -5.632E0 & -1.979E1 & ~1.759E1 &  6.213E1 \\
  8 &  -1.424E0 & -5.082E0 & ~4.517E0 &  1.602E1 \\
  9 &  -2.991E-1& -1.050E0 & ~9.331E-1&  3.303E0 \\
 $\!\!$10 &  -4.869E-1$\!\!$& -1.757E-1$\!\!$&
             ~1.562E-1$\!\!$&  5.557E-1$\!\!$\\
\hline \hline
\multicolumn{1}{||c|}{ } &
\multicolumn{4}{  c||}{ } \\[-0.3cm]
\multicolumn{1}{||c|}{   } &
\multicolumn{4}{  c||}{$N_f = 4$   } \\
\multicolumn{1}{||c|}{ } &
\multicolumn{4}{  c||}{ } \\[-0.3cm] \hline
  & & & \\[-0.3cm]
\multicolumn{1}{||c|}{$l$} &
\multicolumn{1}{c|}{$P_{qq}^{(l)}$} &
\multicolumn{1}{c|}{$P_{qg}^{(l)}$} &
\multicolumn{1}{c|}{$P_{gq}^{(l)}$} &
\multicolumn{1}{c||}{$P_{gg}^{(l)}$} \\
  & & & \\[-0.3cm] \hline \hline
  & & & \\[-0.3cm]
  0 &  ~2.667E0 & -8.000E0 & ~5.333E0 & 2.400E1 \\
  1 &  -1.600E1 & -5.867E1 & ~3.911E1 & 1.227E2 \\
  2 &  -4.953E1 & -1.788E2 & ~1.192E2 & 3.905E2 \\
  3 &  -8.573E1 & -2.859E2 & ~1.906E2 & 6.316E2 \\
  4 &  -7.633E1 & -2.595E2 & ~1.730E2 & 5.831E2 \\
  5 &  -4.649E1 & -1.569E2 & ~1.046E2 & 3.540E2 \\
  6 &  -1.956E1 & -6.688E1 & ~4.459E1 & 1.519E2 \\
  7 &  -6.326E0 & -2.148E1 & ~1.432E1 & 4.882E1 \\
  8 &  -1.546E0 & -5.319E0 & ~3.546E0 & 1.215E1 \\
  9 &  -3.133E-1& -1.063E0 & ~7.086E-1& 2.421E0 \\
 $\!\!$10 &  -4.923E-2$\!\!$& -1.713E-1$\!\!$&
             ~1.142E-1$\!\!$&  3.928E-1$\!\!$\\
\hline \hline
\end{tabular}
\normalsize

\end{center}
\noindent
{\sf {\bf Table~2:}~~The coefficients $\PV^{(l)}$ of the expansion
of $\PV(x,a_s)_{x \rightarrow 0}$ in eq.~(\ref{eqSER}) in terms of
$a_s (a_s \ln^2 x)^l$ for three and four quark flavours.}
\vspace{-6mm}
\end{table}

The leading small-$x$ off--diagonal elements of $\PV (x,a_s)_{x \ra 0}$
are related by
\beq
 P_{qg,l}^{\, x \ra 0}(x)/(T_F N_f) = -P_{gq,l}^{\, x \ra 0}(x)/C_F 
\eeq
 for all $l$.
Also the case of an ${\cal N} = 1$ supersymmetric Yang--Mills field
theory, i.e.\ $C_A = C_F = 1$, $N_f = 1$, and $T_F = 1/2$, has been
considered. The so--called supersymmetric relation
\begin{equation}
 P_{qq,l}(x)+ P_{gq,l}(x)- P_{qg,l}(x)- P_{gg,l}(x) = 0
\end{equation}
is satisfied for the small-$x$ leading terms. In fact even
more restrictive relations are fulfilled in this case,
cf.\ ref.~\cite{BVplb2}.
Finally it should be noted that there is no overlap of the present
small-$x$ resummation with the large-$N_f$ expansion \cite{LNf} of the
all--order splitting function matrix.

\subsection{The unpolarized singlet case}
\label{sect33}

\vspace*{1mm}
\noindent
Unlike the cases discussed in the previous sections, where the leading
small-$x$ singularity in the complex $N$ plane is situated at $N = 0$,
the corresponding poles of the anomalous dimensions for the
unpolarized singlet evolution are located at $N=1$. The all--order
resummation of the most singular contributions as $x\ra 0$, $\gamma_L
(N,a_s)$, to the anomalous dimensions was derived in~\cite{LIPAT}.
It is found as the solution of
\beq
\label{LIP}
 N = 4 C_A a_s \chi[\gamma_L(N,a_s)] \: ,
\eeq
with
\beq
 \chi (\gamma) \equiv 2 \psi (1) -\psi (\gamma ) - \psi (1-\gamma )\: .
\eeq
Here $\psi(z)$ denotes the logarithmic derivative of Euler's $\Gamma
$-function. $\gamma _L $ is a multi-valued function for complex $N$.
The perturbative branch is selected by requiring
\beq
 \gamma_L(N,a_s) \rightarrow \frac{4C_A\, a_s}{N-1}
 ~~{\rm for}~~|N| \rightarrow \infty
\eeq
when solving eq.~(\ref{LIP}). The singularity structure of the solution
in the complex $N$ plane was studied in detail in refs.~\cite{EHW,JBKT}.
Finally the resummed small-$x$ contributions $\GGV _L(N)$ to the
singlet anomalous dimension matrix $\GGV (N) $, related to the splitting
functions in eq.~(\ref{Ps}) by
\beq
 \GGV (N,a_s) = -2 \int_0^1 \! dx \, x^{N-1} \PV (x,a_s) \: ,
\eeq
are in this approximation obtained by
\beq
\label{GAML}
\GGV_L(N, a_S) = -2
 \left( \begin{array}{cc} \! 0 \!       & \! 0 \! \\
                          \! C_F/C_A \! & \! 1 \! \end{array} \right)
 \gamma _L (N, a_s ) \: .
\eeq

The $O(a_s)$ correction $\GGV _{\rm NL}(N,a_s)$ to this most singular
part as $x \ra 0$ have been calculated in ref.~\cite{CH} for the quark
anomalous dimensions $\gamma_{qq,qg}$. In the DIS factorization scheme
the matrix of these next-to-leading small-$x$ contributions is given
by
\bea
\label{GAMNL}
 \lefteqn{ \GGV_{\rm NL}(N, a_s) =
 -2 \left( \begin{array}{cc} \!\! \frac{C_F}{C_A} [ \gamma_{\,\rm NL} -
 \frac{8}{3} a_s T_F ]  & \!\gamma_{\,\rm NL} \! \\
 \!\gamma_{gq,{\rm NL}} & \!\gamma_{gq,{\rm NL}} \! \end{array} \right)
 \! ,} \nonumber \\
  & &
\eea
where $\gamma_{\,\rm NL}$ is used as an abbreviation for the
function $\gamma_{\,\rm NL}^{\rm DIS}(N,a_s)$. It can be
recursively expressed by $\gamma_L(N,a_s)$ and reads
\bea
 \lefteqn{
 \gamma _{\,\rm NL}^{\rm DIS}(N, a_s) = } \\
 & & \hspace*{-6.5mm} 24 T_F a_s
  \frac{2 + 3\gamma _L - 3\gamma _L^2}{3 - 2\gamma _L}
  \frac{[B(1- \gamma _L, 1+ \gamma _L)]^3}
  {B(2+ 2\gamma _L, 2- 2\gamma _L)} R( \gamma _L)
  \nonumber
\eea
with $B(x,y)$ denoting the Beta function and
\bea
 \lefteqn{
 R(\gamma ) =  \left[ \frac{\Gamma (1- \gamma )\chi (\gamma )}
 {\Gamma (1+ \gamma ) \{ -\gamma \chi ^{\prime}(\gamma )\}}\right]^{1/2}
 } \\
 & & \hspace{3mm}\exp \left[ \gamma \psi (1) + \int_0^{\gamma }\! d\zeta
 \, \frac{\psi ^{\prime}(1) - \psi ^{\prime}(1 - \zeta )}{\chi (\zeta )}
 \right]. \nonumber
\eea
The calculation of the yet unknown gluonic entries $\gamma_{gq,\rm NL}$
and $\gamma_{gg, \rm NL}$ in eq.~(\ref{GAMNL}) is in progress
\cite{GGNL,CC}. A first contribution to $\gamma_{gg, \rm NL}$
$\propto N_f$ has been determined
recently~\cite{CC} in the $Q_0$-scheme~\cite{JBDUR,Q0S},
which has been introduced in the framework of $k_{\perp}$-factorization.

Both $\GGV_L$ and $\GGV_{\,\rm NL}$ can be represented by infinite
series in $a_s$. The analytic expressions are straightforwardly
obtained but are rather lengthy. The numerical size of the resulting
coefficients $A_l$ and $B_l$ in the DIS scheme is illustrated in
Table~3. The matrix $ \PV(x, a_s)_{\rm res}$ of the splitting functions
including
the small--$x$
resummed terms (cf.\ Section 4.4)
beyond the complete  LO and NLO matrices from the
fixed--order calculations reads:
\bea
\label{PSres}
 \lefteqn{ \PV(x, a_s)_{\rm res}^{\rm DIS} =
 a_s \PV_{0}(x) + a_s^2 \PV_{1}(x)^{\rm DIS} } \nonumber \\
 & & + \sum_{l=2}^{\infty} A_l\,  a_{s}^{l+1} \frac{1}{x}
   \ln^{l} \left(\frac{1}{x}\right)
   \left( \begin{array}{cc} \! 0        & \! 0 \! \\
                            \! C_F/C_A  & \! 1 \! \end{array}\right) \\
 & & + \sum_{l=1}^{\infty} B_l\,  a_{s}^{l+2} \frac{1}{x}
   \ln^{l} \left(\frac{1}{x}\right)
   \left( \begin{array}{cc} \! C_F/C_A & 1 \! \\
                                  \! 0 & 0 \! \end{array} \right) .
 \nonumber
\eea
In the subsequent numerical treatment we will use the labels {\it 
`Lx'\/} for results obtained with the leading series $(A_l)$, and {\it 
`NLx'\/} in the cases where the $B_l$-terms in eq.~(\ref{PSres}) have 
been taken into account additionally.

\begin{table}[t]
\vspace{-2mm}
\begin{center}
\small
\begin{tabular}{||r|l|l||}
\hline \hline
 & & \\[-0.3cm]
\multicolumn{1}{||c|}{$l$} &
\multicolumn{1}{c|} {$A_l$} &
\multicolumn{1}{c||}{$B_l$} \\
 & & \\[-0.3cm] \hline \hline
 & & \\[-0.3cm]
      0 &  ~1.200E1   & $N_f \cdot $ ~3.467E1   \\
      1 &  ~0.000E0   & $N_f \cdot $ ~4.415E2   \\
      2 &  ~0.000E0   & $N_f \cdot $ ~9.528E3   \\
      3 &  ~8.308E3   & $N_f \cdot $ ~6.877E4   \\
      4 &  ~0.000E0   & $N_f \cdot $ ~4.040E5   \\
      5 &  ~5.160E4   & $N_f \cdot $ ~3.411E6   \\
      6 &  ~8.629E5   & $N_f \cdot $ ~1.293E7   \\
      7 &  ~1.721E5   & $N_f \cdot $ ~5.518E7   \\
      8 &  ~5.104E6   & $N_f \cdot $ ~2.601E8   \\
      9 &  ~2.879E7   & $N_f \cdot $ ~7.086E8   \\
     10 &  ~1.433E7   & $N_f \cdot $ ~2.343E9   \\
\hline \hline
\end{tabular}
\normalsize
\end{center}

\noindent
{\sf {\bf Table~3:}~~The coefficients $A_l$ and $B_l$ of the expansion
for the small-$x$ resummed splitting functions in the DIS scheme, see
eq.~(\ref{PSres}). For completeness also $A_0$, $A_1$ and $B_0$ are
given. For a more precise  representation and higher-$l$ terms, cf.\
refs.~\cite{BF,BRVup}.}
\vspace{-5mm}
\end{table}

Finally we list the presently available predictions from the small-$x$
resummations for the most singular contributions $P_{ij,2}^{\, x \ra 0}$
of the NNLO splitting functions. There are no $a_s^3 \ln^2 x$ terms,
due to the matrix structure of eq.~(\ref{GAML}) and the vanishing of
$A_2$ in eq.\ (\ref{PSres}). For the quark splitting functions \cite{CH}
the $a_s^3 \ln x$ terms read in the DIS scheme
\bea
\lefteqn{ P_{qq,2}^{\, x \ra 0}(x)^{\rm DIS} = \,
  \left[ \frac{568}{9} - \frac{8}{3} \pi^2 \right]
  C_F\, C_A \frac{1}{x} \ln \left( \frac{1}{x}\right) \:. }
 \nonumber \\ & &
\eea
The corresponding \MSbar\ results are given by
\beq
 P_{qq,2}^{\, x \ra 0}(x)^{\overline{\rm MS}} = \,
  \frac{224}{9} C_F\, C_A \frac{1}{x} \ln \left( \frac{1}{x}\right) \: ,
\eeq
and 
\beq
 P_{qg,2}^{\, x \ra 0}(x) =
      (C_A/C_F) P_{qq,2}^{\, x \ra 0}(x)
\eeq
in both schemes.
Unlike the cases discussed in the previous sections, the coefficient
functions contain terms as singular as the splitting functions in the
$\overline{\rm MS}$ scheme.

\noindent
\subsection{Less singular contributions}
\label{sect34}

\vspace*{1mm}
\noindent
The terms in the splitting functions $P_{ij}$ and $P^{\pm}$, which are
less singular by one (or more) powers of $\ln (1/x)$ as $x \ra 0 $
than the leading contributions discussed in the previous sections, are
presently unknown in almost all cases.
Such subleading contributions, however, can potentially prove to be as
important as the leading terms, since the splitting functions and
coefficient functions enter observable quantities always via Mellin
convolutions with the parton distributions.

This situation can be illustrated by a simple example, cf.\ ref.~\cite
{JBDUR}. Consider the lowest--order gluonic contribution to the
longitudinal structure function $F_L(x,Q^2)$, given by
\bea
\lefteqn{ F_L^{\, g}(x,Q^2) = } \\
 & & 8\, a_s \sum_q e_q^2 \int_x^1 \!\frac{dy}{y} \, y^2(1 - y) \,
     \frac{x}{y} g(x/y,Q^2)\: . \nonumber
\eea
If one replaces the term $1 - y$ originating from the coefficient
function by its small-$y$ approximation 1 for small values of $x$, the
result for $F_L^{\, g}$ changes by a factor of about 4 for typical
parametrizations of the gluon density! Due to the Mellin convolution and
the fact that $g(x)$ becomes very large as $x \ra 0$, the coefficient
function at medium and large $y$ contributes essentially. 
On the other hand the coefficient function in the range $y \gsim x$, 
where the small-$y$ approximation is justified, samples $g(x/y \lsim 1)$
which is however small. Similar observations can be made for other 
convolutions considered in the previous sections as well.

The non--singlet `--' and the unpolarized singlet splitting functions
are constrained by conservation laws, see eq.~(\ref{conserv}). The
resummed contributions discussed in Sections~3.1 -- 3.3 do not obey
these constraints, however, less singular terms restore these sum 
rules. Also for the cases in which the anomalous dimensions are not 
subject to such constraints, less singular terms with sizeable 
coefficients exist for example in NLO, e.g.\ eq.~(\ref{NSsub}).

In order to evaluate the possible impact of such terms, their numerical
coefficients have to be estimated. At present the only source of
information are the fully known LO and NLO splitting functions. The
dominant and subdominant terms as $x \ra 0 $ for the NLO anomalous
dimensions are recalled in eqs.~(\ref{NSsub})--(\ref{DISsub}).
The results are presented in the general form, as well as, for easier
comparison of the numerical size of the coefficients, inserting the
number of active flavours as used in the numerical applications in
Section~4.  In the non--singlet cases one finds in the \MSbar\ scheme
%
\bea
\label{NSsub}
 \gamma_1^+(N)_{x\ra 0} &=& -\frac{128}{9N^3}\:\, + \frac{400 - 32 N_f}
    {9N^{2}} \nonumber \\[1mm]
 &\hspace*{-2mm}\stackrel{N_f=4}{=}\hspace*{-2mm}&
 -\frac{14.22}{N^3} + \frac{30.22}{N^2} \: , \\[1mm]
 \gamma_1^-(N)_{x\ra 0} &=& -\frac{64}{3N^3}\:\,  + \frac{464 - 32 N_f}
    {9N^{2}} \nonumber \\[1mm]
 &\hspace*{-2mm}\stackrel{N_f=4}{=}\hspace*{-2mm}&
 -\frac{21.33}{N^3} + \frac{37.33}{N^2} \: . \nonumber
\eea
Note that the subleading terms are roughly of the same size in both
cases, despite only one of the combinations being constrained by a sum
rule.
The corresponding results for the polarized singlet case, also in the
\MSbar\ scheme, are given by
\bea
 \gamma_{qq,1}^{\rm pol}(N)_{x\ra 0} &=& \frac{-64 + 64 N_f}{3N^3} +
    \frac{464-128 N_f}{9N^2} \nonumber \\[1mm]
 &\hspace*{-2.2mm}\stackrel{N_f=3}{=}\hspace*{-2mm}&
    +\frac{42.67}{N^3} + \frac{8.889}{N^2} \: , \nonumber \\[1mm]
 \gamma_{qg,1}^{\rm pol}(N)_{x\ra 0} &=& +\frac{176 N_f}{3N^3} -
    \frac{24 N_f} {N^2} \nonumber \\[1mm]
 &\hspace*{-2.2mm}\stackrel{N_f=3}{=}\hspace*{-2mm}&
    +\frac{176.0}{N^3} - \frac{72.00}{N^2} \: , \\[1mm]
 \gamma_{gq,1}^{\rm pol}(N)_{x\ra 0} &=& -\frac{1408}{9N^3}\: +
    \frac{896}{9N^2} \nonumber \\[1mm]
 &\hspace*{-2.2mm}\stackrel{N_f=3}{=}\hspace*{-2mm}&
    -\frac{156.4}{N^3} + \frac{99.56}{N^2} \: , \nonumber \\[1mm]
 \gamma_{gg,1}^{\rm pol}(N)_{x\ra 0} &=& \!\!\!\!\!\frac{-1728 + 64 N_f}
    {3N^3} \! + \!\frac{2088 - 208 N_f}{3N^2} \nonumber \\[1mm]
 &\hspace*{-2.2mm}\stackrel{N_f=3}{=}\hspace*{-2mm}&
    -\frac{512.0}{N^3} + \frac{488.0}{N^2} \: . \nonumber
\eea

\vspace{\fill}
\noindent
The first two terms of the unpolarized singlet anomalous dimensions
expanded at $N=1$ read in the DIS scheme
\bea
\label{DISsub}
 \gamma_{qq,1}^{\rm DIS}(N)_{x\ra 0} \!\! &=& \!\!\!\!\frac{-832 N_f}
    {27(N-1)} + 2.56 + 100.83 N_f
    \nonumber \\[1mm]
 &\hspace*{-3.5mm}\stackrel{N_f=4}{=}\hspace*{-2mm}&
    -\frac{123.3}{N-1} + 405.9 \: , \nonumber \\[1mm]
 \gamma_{qg,1}^{\rm DIS}(N)_{x\ra 0} \!\! &=& \!\!\!\!\frac{-208 N_f}
     {3(N-1)} + 218.67 N_f - 1.78 N_f^2
    \nonumber \\[1mm]
 &\hspace*{-3.5mm}\stackrel{N_f=4}{=}\hspace*{-2mm}&
    -\frac{277.3}{N-1} + 846.2 \: , \\[1mm]
 \gamma_{gq,1}^{\rm DIS}(N)_{x\ra 0} \!\! &=& \!\!\!\!\!\!\frac{-864 + 
    832 N_f} {27(N-1)} \! - \! 185.8 \! - \! 66.94 N_f
    \nonumber \\[1mm]
 &\hspace*{-3.5mm}\stackrel{N_f=4}{=}\hspace*{-2mm}&
    +\frac{91.26}{N-1} - 453.5 \: , \nonumber
\eea
\newpage
\bea
 \gamma_{gg,1}^{\rm DIS}(N)_{x\ra 0} \!\! &=& \!\!\!\!\frac{184 N_f}
    {3(N-1)} - 629.8 - 93.17 N_f \nonumber \\[-2mm]
 & & \hspace*{25mm} \mbox{}  + 0.889 N_f^2
    \nonumber \\
 &\hspace*{-3.5mm}\stackrel{N_f=4}{=}\hspace*{-2mm}&
    +\frac{245.3}{N-1} - 988.3 \: . \nonumber
\eea
One notices that the subleading terms occur in general with signs 
opposite to those of the dominant ones. Their prefactors are of the 
same order, but in most cases a factor of about 2 to 4 larger.

Thus introducing subleading terms with prefactors up to two times larger
than those of the leading terms appears to yield reasonable and
conservative estimates for the possible impact of subleading terms.
The following modifications of the resummed anomalous dimensions 
$\Gamma(N,\alpha_s)$ have accordingly been studied within  
refs.~\cite{BVplb1,BVcrac,BVplb2,EHW,BRVup}:
\begin{equation}
\label{xxx}
\begin{array}{cl}
{\rm A:} & \Gamma(N, \alpha_s)
\rightarrow  \Gamma(N, \alpha_s) - \Gamma(1, \alpha_s)
\\
{\rm B:} & \Gamma(N, \alpha_s)
\rightarrow  \Gamma(N, \alpha_s)(1 - N)
\\
{\rm C:} & \Gamma(N, \alpha_s)
\rightarrow  \Gamma(N, \alpha_s)
(1 - 2N + N^2)
\\
{\rm D:} & \Gamma(N, \alpha_s)
\rightarrow  \Gamma(N, \alpha_s)(1 - 2N + N^3) \:\: ,
\end{array}
\end{equation}
where the replacement $ N \rightarrow N-1 $ is understood for the case 
of Section 3.3.

Let us finally discuss also the case of the `--' non--singlet evolution 
in QED.
For the evolution kernel also the terms of $O(\alpha^2 \ln x )$ were 
calculated for $e^+e^-$ annihilation in the on--mass--shell scheme 
(OMS) in ref.~\cite{BBN}. All contributions but those due to the vacuum 
polarization diagram cancel in this order. One obtains
\begin{eqnarray}
\label{KQED} \left.
K_{1,x \rightarrow 0}^{-, {\rm QED}}(x,a) \right|_{\rm OMS}
 &\!\!\! =\!\!\! & -6 a^2 \left [ \ln^2 x + \frac{4}{9} \ln x \right ] 
 \nonumber\\
&\!\!\!  \leftrightarrow \!\!\! & -12 \frac{a^2}{N^3} 
 \left [ 1 - \frac{2}{9} N \right ].
\end{eqnarray}
Unlike for most of the examples discussed above, in this particular
case the term being suppressed by one order in $\ln x $ has thus
a smaller coefficient than the leading singular contribution.

\section{Numerical consequences}
\label{sect4}
\subsection{The unpolarized non-singlet case}
\label{sect41}

\vspace*{1mm}
\noindent
The evolution equations (\ref{evolf}) for the non--singlet combinations
of structure functions can be solved analytically in Mellin-$N$ space.
Taking into account the resummed kernels $K_{x \rightarrow 0}^{\pm}$ of
eq.~(\ref{Kpm}) in addition to the full splitting functions up to NLO,
the solution can be written as \cite{BVplb1,BVcrac}
\bea
\label{Sns}
 \lefteqn{ F^{\pm }(N,a_s) = F^{\pm}(N,a_0) \left( \frac{a_s}{a_0}
     \right)^{\gamma_{0}(N)/2\beta_0} } \nonumber \\
 & & \hspace*{-4mm}\times \left \{ \exp \left[ \frac{1}{2\beta_0}
     \int_{a_{0}}^{a_s} \!\! da \, \frac{1}{a^2} \Gamma ^{\pm}(N,a_s)
     \right] + \frac{a_s - a_0}{2\beta_{0}} \right. \nonumber \\
 & & \left. \cdot \left[ \tilde{\gamma}_1^{\pm }(N) - \frac{\beta_1}
     {\beta_{0}} \gamma_{0}(N) + 2\beta_0 \hat{c}^{\pm}_{i,1}(N)
     \right] \right \},
\eea
with
\bea
\label{zzz}
\gamma_{\, l}(N)
 &\! =\! & -2 \int_0^1 \! dx \, x^{N-1} P_l(x)
 \: , \nonumber \\
\hat{c}_{i,l}(N)
 &\! =\! & \int_0^1 \! dx \, x^{N-1} c_{i,l}(x) \: ,
\eea
and $a_0 = a_s(Q^2_0)$.
Here $ \tilde{\gamma}_1^{\pm }(N) $ denotes the two--loop anomalous
dimension $ \gamma_1^{\pm }(N) $ with the leading $ 1/N^3 $ term 
subtracted. The effect of this term is included to all orders in the
exponential factor, which in turn is obtained from eq.~(\ref{Gns}) by
removing the LO contribution included in $\gamma_{0}(N)$:
\bea
\label{SOL1}
 \Gamma ^{\pm}(N,a_s)
 &\!\! =\!\! & \GA _{x \ra 0}^{\pm}(N,a_s) - \frac{a_s}{N}
   \lim_{N \ra 0} \, [N\gamma_{0}(N)] \nonumber \\
 &\!\! =\!\! & \GA _{x \ra 0}^{\pm}(N,a_s) + a_s \frac{4C_F}{N} \: .
\eea
The NLO evolution of $F^{\pm}(N,a_s)$ can be recovered from eq.\
(\ref{Sns}) by expanding the exponential to first order in $a_s$ and 
$a_0$. The inverse Mellin transformation of the final results back to 
$x$-space is performed by a numerical integral in the complex
$N$-plane, see e.g.\ ref.~\cite{GRV90}.

The remaining quadrature in (\ref{Sns}) can be performed analytically
for the `+'-case \cite{BVcrac}, and has to be done numerically for the
`--'-combinations involving the parabolic cylinder function $ D_p(z\!
=\! N/\sqrt{2N_{c}a_s}) $. Additional information can be obtained by
expanding the resummed kernels $ \Gamma ^{+}(N,a_s) $ and $ \Gamma
^{-} (N,a_s) $ in $a_s$, in the latter case using the asymptotic
expansion of $D_{p}(z)$ \cite{RYGRA}.

The evolution of the `--'-combination
\bea
\label{F3N}
 \lefteqn{ xF_{3}^{\, N}(x,Q_0^2) \equiv \frac{1}{2} \!\left[ xF_{3}^{\,
    \nu N} (x,Q_0^2) + xF_{3}^{\, \bar{\nu} N}(x,Q_0^2) \right ] }
    \nonumber \\
 & & = c_{F_3}^-(x,Q^2_0) \otimes [xu_v + xd_v](x,Q_0^2)
\eea
for an isoscalar target $N$ and the  `+'-combination
\bea
\label{F2pn}
  \lefteqn{ F_{2}^{\, ep}(x,Q_0^2) - F_{2}^{\, en}(x,Q_0^2)
  = } \\
 & & \hspace{-7mm} c_{F_2}^+(x,Q^2_0) \otimes \frac{1}{3}
  \!\left[ xu_v - xd_v - 2(x\bar{d}-x\bar{u})\right]\! (x,Q^2_0)
  \nonumber
\eea
have been investigated in refs.~\cite{BVplb1,BVcrac}. As in all other
numerical examples displayed below, the reference scale for the
evolution (\ref{Sns}) is chosen as $Q_{0}^{2} = 4 \mbox{ GeV}^2$, and
the same input parameters are employed for the NLO and the resummed
calculations. In the present case, the initial parton distributions
have been  adopted from the MRS(A) global fit~\cite{MRSA} together with 
the value of the QCD scale parameter, $\Lambda_{\overline{\rm MS}}
(N_f =4) = 230 \MeV$.

\begin{figure}[htb]
\vspace*{-15mm}
\begin{center}
\mbox{\hspace*{-4mm}\epsfig{file=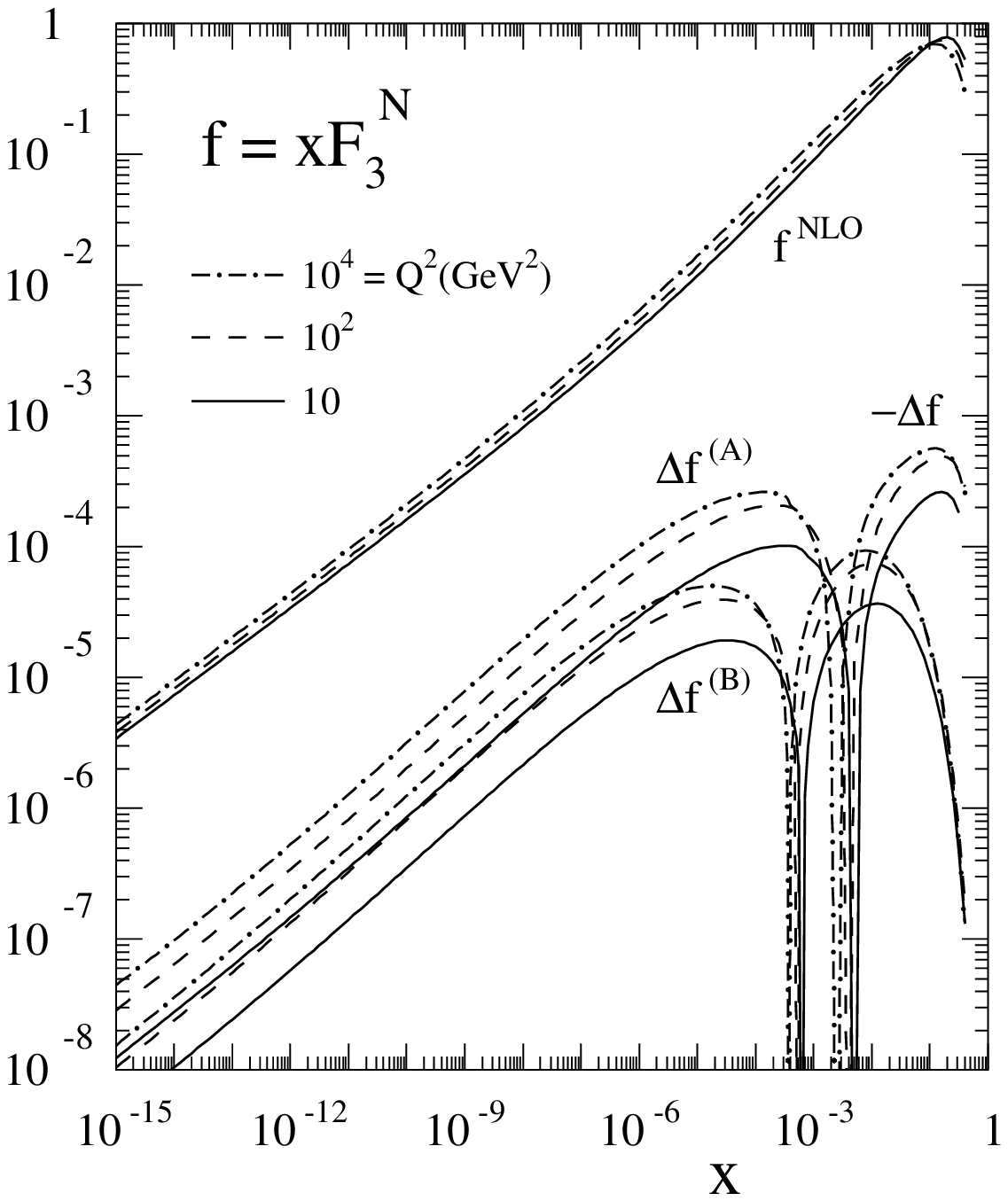,width=8.4cm}}
\vspace*{-10mm}
\end{center}
{\sf {\bf Figure 1:}~~The small-$x$ $Q^2$-evolution of the non--singlet
 isoscalar structure function $xF_3^{\, N}$ in NLO and the absolute
 corrections to these results due to the resummed kernels of Section
 3.1. `(A)' and `(B)' denote two prescriptions for implementing fermion
 number conservation, see eq.~(\protect\ref{xxx}).}
\vspace{-1mm}
\end{figure}
\begin{figure}[thb]
\vspace*{-10mm}
\begin{center}
\mbox{\hspace*{-4mm}\epsfig{file=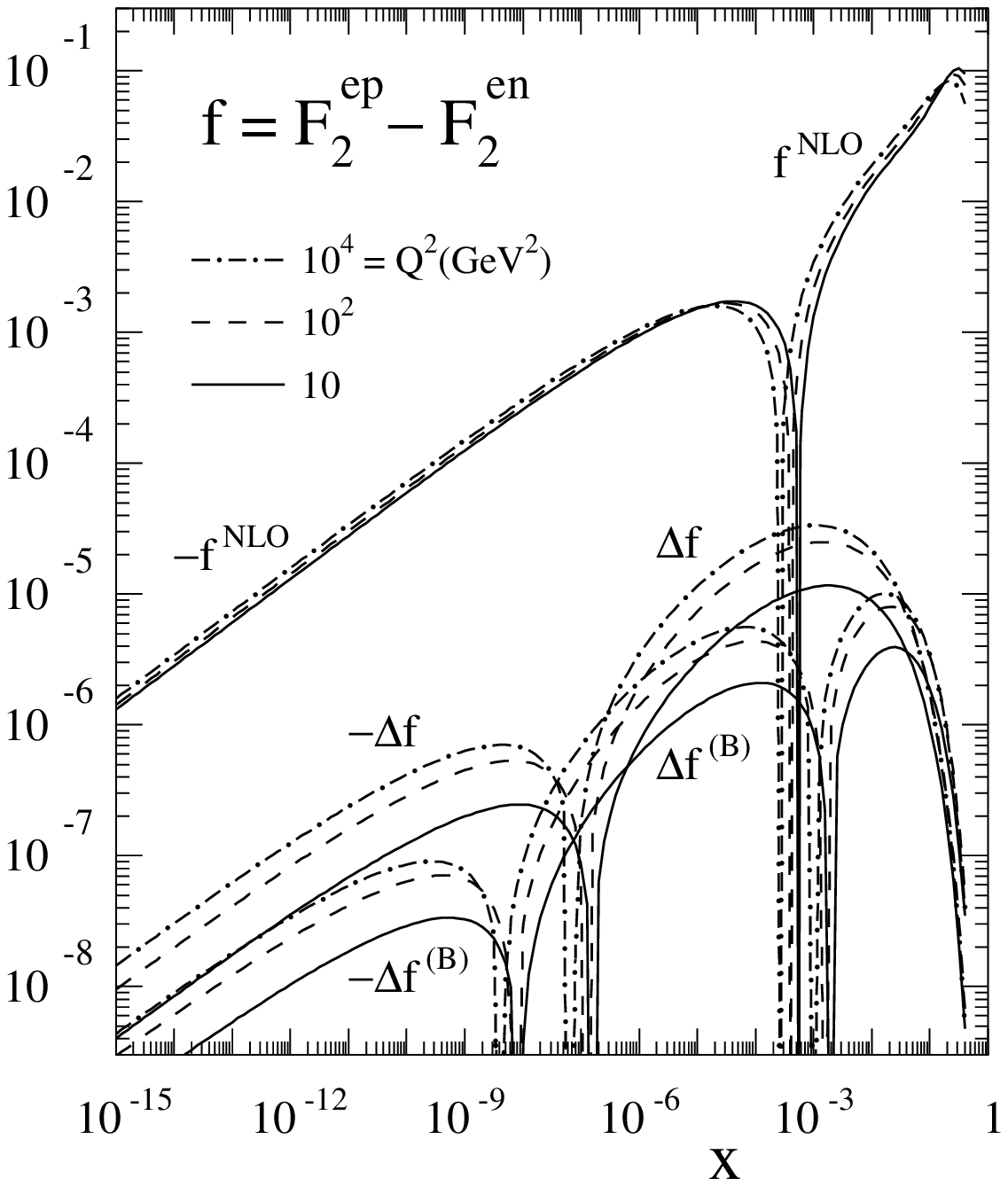,width=8.2cm}}
\vspace*{-11mm}
\end{center}
{\sf {\bf Figure 2:}~~The same as in Figure 1, but for the structure
 function combination $F_2^{\, ep} - F_2^{\, en}$. Instead of the
 prescription `(A)', the result without any subleading terms is shown
 for this `+'-case.}
\vspace{-7mm}
\end{figure}

\noindent
The small-$x$ behaviour of the most relevant initial distributions is
given by $ xu_v(x,Q_{0}^{2})\sim x^{0.54} $ and $ xd_v(x,Q_{0}^{2})\sim
x^{0.33}$ \cite{MRSA}. Hence these distributions are rather `steep':
their rightmost singularities in the complex $N$-plane lie about 0.5
units or more to the right of the leading singularity of the
non--singlet anomalous dimensions at $N=0$.

In Figures~1 and 2 the NLO results for $xF_3^N$ and $F_{2}^{ep} -
F_{2}^{en}$ are shown, together with the corresponding resummation
corrections, down to $x$ as low as $x = 10^{-15}$. Even at these 
extremely small values of $x$, the effect of the resummed anomalous 
dimensions stays at the level of 1\% or below. This is in striking 
contrast to the expectation of ref.~\cite{EMR}, where corrections of 
up to factors of 10 in the HERA kinematical regime were anticipated.
Moreover the resummation effects remain very sensitive to presently
unknown terms less singular as $x \ra 0 $ in the splitting functions.
This is illustrated by the impact of the prescription $(B)$ of
eq.~(\ref{xxx}), which removes as much as about two thirds of the
resummation effect even at asymptotically low $x$.

Another interesting issue is the $a_s$-expansion of the resummed kernel
$\GA _{x \ra 0}^{\pm} $ (\ref{Gns}) mentioned before.  In Figure~3 the
resummation corrections with $\Gamma ^{\pm}(N,a_s)/a_s$ in 
eq.~(\ref{Sns}) expanded to order $l$ are compared to the full results 
for the cases with no subtraction at one typical value of $Q^2$.

\begin{figure}[htb]
\vspace{-16mm}
\begin{center}
\mbox{\hspace*{-5mm}\epsfig{file=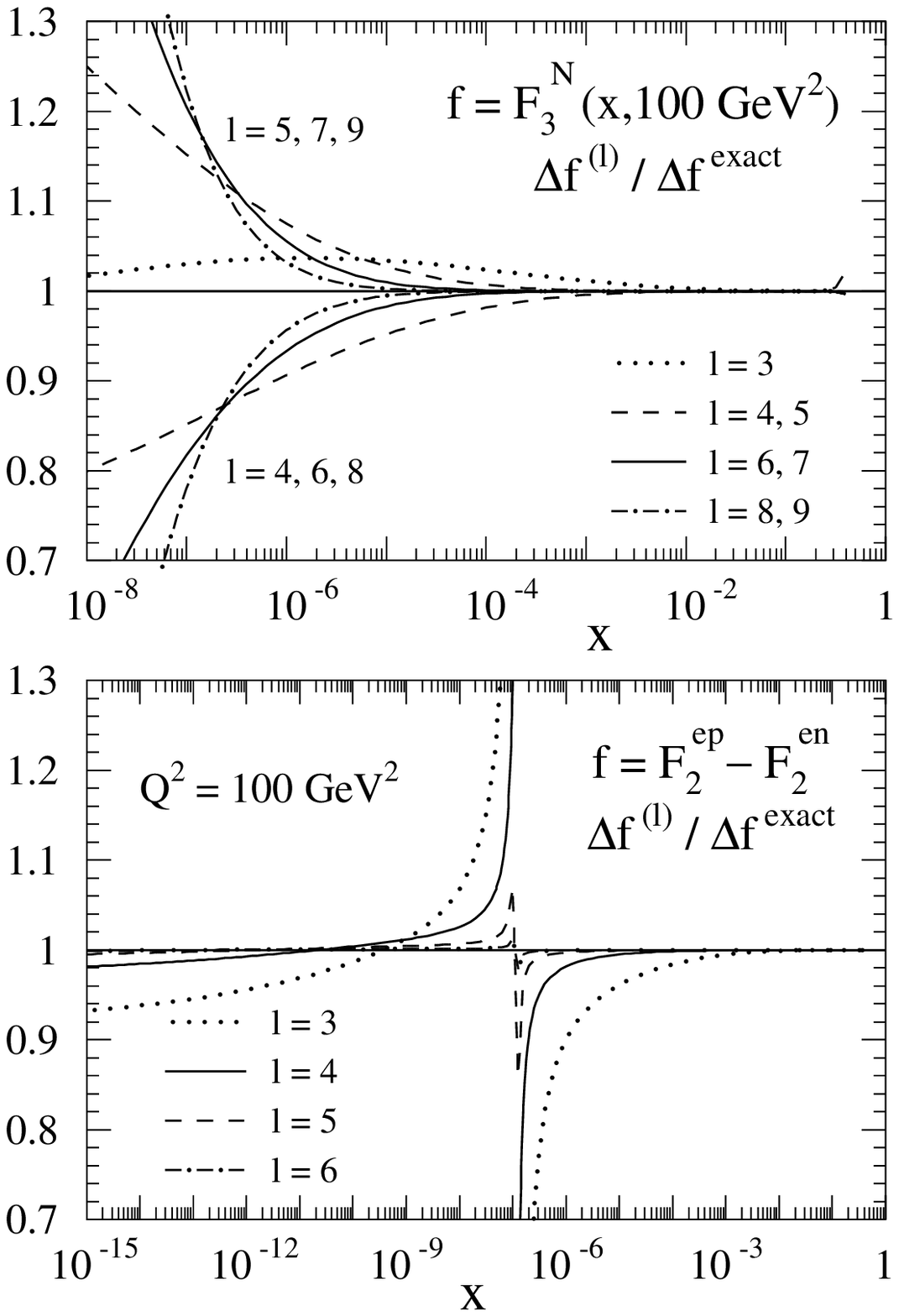,width=8.4cm}}
\vspace*{-12mm}
\end{center}
{\sf {\bf Figure 3:}~~The ratio of the resummation corrections with
 the resummed anomalous dimensions $\GA _{x \ra 0}^{\pm} $ expanded at
 order $a_s^{l+1}$ to the complete effects for the `--'-quantity
 $F_3^{\, N}$ (upper part) and the `+'-combination $F_2^{\, ep} -
 F_2^{\, en}$ (lower part).}
\vspace{-7mm}
\end{figure}

\noindent
The asymptotic series for the `--'-case leads to a good approximation
at $x \gsim 10^{-6} $, but starts to diverge below.
In the
`+'-case, on the other hand, the Taylor
series   converges  in the
whole
$x$-range considered. In both cases the next two terms beyond NLO,
$l=3$, contribute more than 90\% of the final resummation effect, again
even down to $x=10^{-15}$. The reduced stability for $F_2^{\, ep} -
F_2^{\, en}$ around $x= 10^{-7}$ is immaterial, since $\Delta f$
changes sign here.

\subsection{The polarized non--singlet case}
\label{sect42}

\vspace*{1mm}
\noindent
The solution of the evolution equations proceeds in the same way as in
the previous section, see eqs.~(\ref{Sns})--(\ref{SOL1}). The present
case is however practically even 
more interesting. Firstly the non--singlet
structure functions are, unlike in the unpolarized case, not a priori
suppressed versus their singlet counterparts at very low $x$. Secondly
the shapes of the polarized parton densities are not  well
established yet~\cite{polrev} by  the
experimental results. 
This is illustrated
in Figure~4, where the small-$x$ extrapolations of the structure 
function $g_{1}^{\, ep} - g_{1}^{\, en}$ (see eq.~(\ref{g1pn}) below) 
are compared for two choices of the initial distributions.

\begin{figure}[htb]
\vspace{-15mm}
\begin{center}
\mbox{\hspace*{-3mm}\epsfig{file=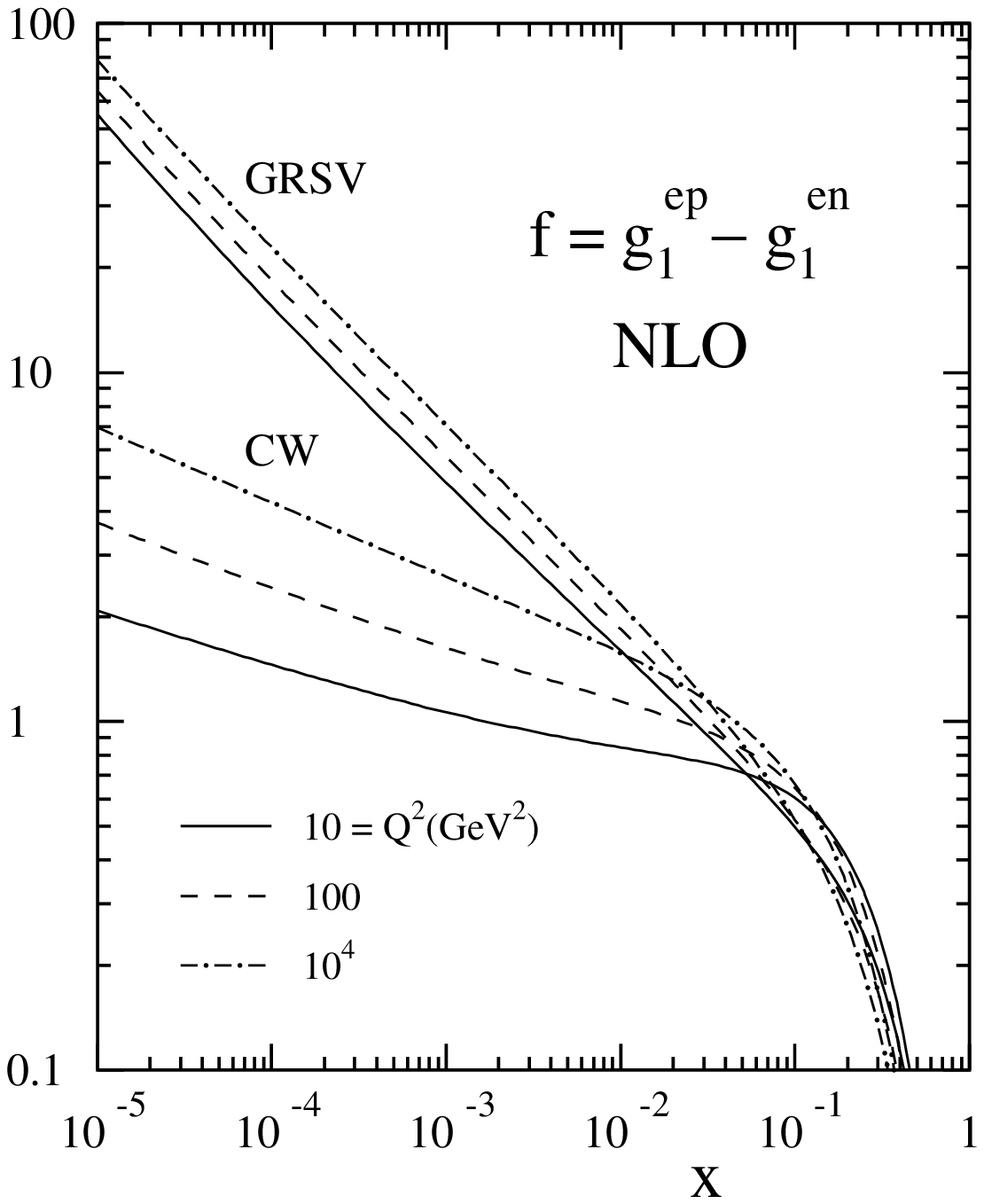,width=7.8cm}}
\vspace{-10mm}
\end{center}
{\sf {\bf Figure 4:}~~The NLO evolution of the polarized non--singlet
 structure function combination $g_{1}^{\, ep} - g_{1}^{\, en}$
 for the input densities from refs.~\cite{CW} and \cite {GRSV}.}
\vspace{-10mm}
\end{figure}

At small $x$ the valence quark densities of CW \cite{CW} are relatively
flat: $\Delta u_v,\, \Delta d_v (x,10\, {\rm GeV}^2) \sim x^{-0.17, \,
+0.29}$. The more recent distributions of GRSV \cite{GRSV} are on the
other hand approximately as steep as the unpolarized distributions, 
$\Delta u_v (x,Q_0^2) \sim x^{-0.28}$ and  $\Delta d_v (x,Q_0^2) 
\sim x^{-0.67}$.  For the evolution of the former (latter) input 
$\Lambda_{\overline {\rm MS}}(N_f\! =\! 4) = 230\: (200) \mbox{ MeV} $ 
is employed in this section.

The effect of the resummed anomalous dimensions $\GA ^{\pm}_{x \ra 0}$
(\ref{Gns}) has been investigated in refs.~\cite{BVplb1,BVcrac} for
the `--'-combination
\bea
\label{g1pn}
 \lefteqn{ g_{1}^{\, ep}(x,Q^2_0) - g_{1}^{\, en}(x,Q^2_0) = } \\
  & & \hspace*{-7mm} c^-_{g_1}(x,Q^2_0) \otimes \frac{1}{6}\!
  \left[ \Delta u_v\! -\! \Delta d_v + 2(\Delta \bar{u}\! -\!
  \Delta \bar{d}) \right] \! (x,Q^2_0) \, . \nonumber
\eea
Here also the `+'-case of the $\gamma Z$-interference structure
function \cite{BKplb} is considered:
\bea
\lefteqn{ g_{5,\gamma Z}^{\, ep}(x,Q^2_0) =
c^+_{g_5}(x,Q^2_0) \otimes \frac{1}{4}\! \left[ \Delta u_v\! +
 \Delta d_v\! \right]\! (x,Q^2_0) \, . \nonumber } \\
  & &
\eea

\vspace{-1mm}
\noindent
The resummation corrections for $g_{1}^{\, ep} - g_{1}^{\, en}$ and
$g_{5,\gamma Z}^{\, ep}$ are shown in Figures 5 and 6, respectively.
\begin{figure}[h]
\vspace*{-24mm}
\begin{center}
\mbox{\hspace{-3mm}\epsfig{file=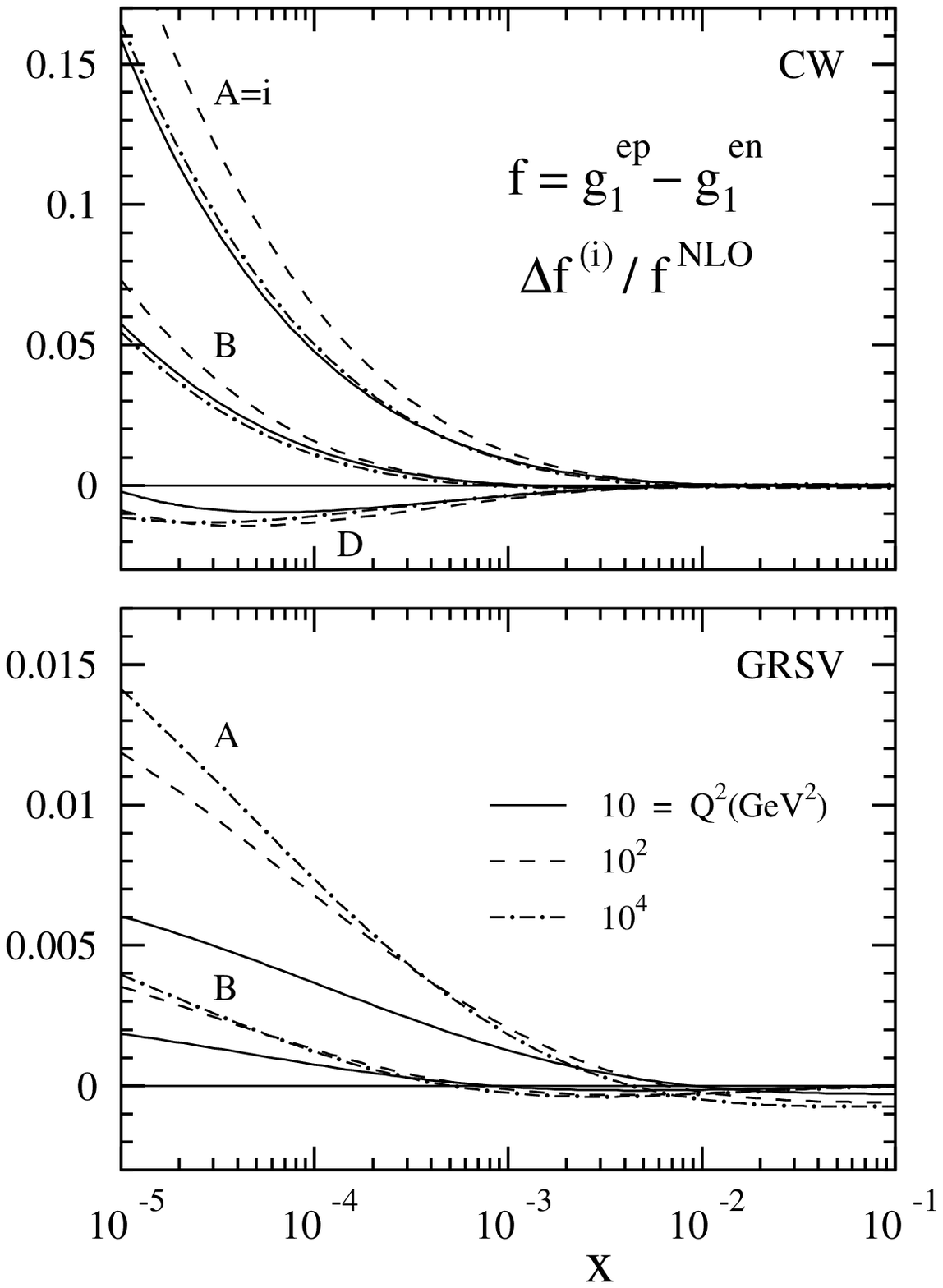,width=8.2cm}}
\vspace*{-10mm}
\end{center}
{\sf {\bf Figure 5:}~~Relative corrections to the NLO small-$x$
 $Q^2$-evolution of the `-'-combination $g_1^{\, ep} - g_1^{\, en}$ due
 to the resummed kernel of Section 3.1 for the initial distributions
 of refs.~\cite{CW} and \cite{GRSV}.
 `A', `B', and `D' denote different prescriptions for implementing
 fermion number conservation, see eq.~(\ref{xxx}).}
\end{figure}

\begin{figure}[htb]
\vspace*{-9mm}
\begin{center}
\mbox{\hspace*{-4mm}\epsfig{file=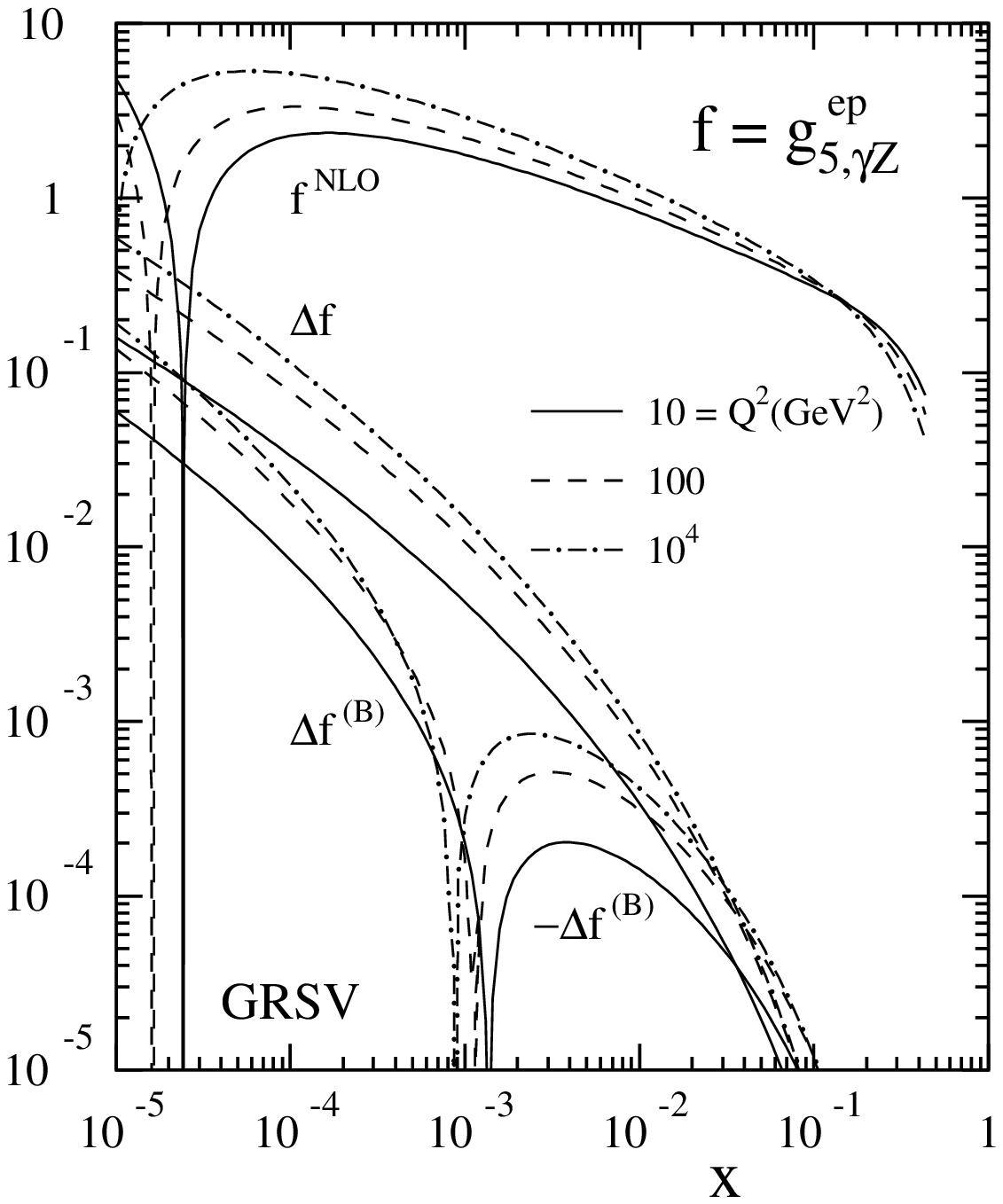,width=8.4cm}}
\vspace{-11mm}
\end{center}
{\sf {\bf Figure 6:}~~The small-$x$ evolution of the non--singlet
 interference structure function $g^{\, ep}_{5,\gamma Z}$ in NLO and
 the  resummation corrections to these results. The possible
 importance of less singular terms in the higher--order splitting
 functions is illustrated by the prescription `(B)' for this `+'-case.}
\vspace{-9mm}
\end{figure}
For the relatively flat CW input \cite{CW}, the resummation effect on
$g_1^{\, ep} - g_1^{\, en}$ reaches
 15\% at $x = 10^{-5}$.
However, in the restricted kinematical range accessible in possible
future polarized electron--polarized proton collider experiments at 
HERA~\cite{JBkin}, it amounts to only  1\% or less. For the steeper
GRSV initial distributions \cite{GRSV}, the effect is of order 1\% or 
smaller in the whole $x$ range, as in the unpolarized cases considered 
above. Hence also here the results do not come up to previous 
expectations  of huge corrections up to factors of 10 or larger 
\cite{BER1} in the HERA range. The outcome is very similar for
$g_{5,\gamma Z}^{\, ep}$, see Figure~6.

As for the unpolarized non--singlet structure functions, the resummation
results also in the polarized cases are not stable against possible
subleading contributions in the higher--order anomalous dimensions.
With respect to the $a_s$ expansions of the kernels $\GA ^{\pm}_{x \ra
0}$ (\ref{Gns}) the situation is also similar to the unpolarized cases.

\subsection{The polarized singlet case}
\label{sect43}

\vspace*{1mm}
\noindent
The solution for the singlet part of the evolution equation 
(\ref{evol2}) cannot be given in a closed form beyond leading order. 
This difference to the non--singlet case is due to the 
non--commutativity of the splitting functions matrices $\PV_l(x)$ in 
eq.~(\ref{Ps}) for different orders in the strong coupling $a_s$. Thus 
the solution has to be written down as a power series in $a_s$, in 
$N$-space resulting in
\bea
\label{Ssi}
 \qV (N,a_s) &\!\!\! = \!\!\! & 
 \Big [ 1 + \sum_{l=1} a_s^l \UV_l (N)\Big]
 \left( \frac{a_s}{a_0} \right)^{{\small \gV}_0 (N)/2\beta_0}
 \nonumber \\
 & & \Big[ 1 + \sum_{l=1} a_0^l \UV_l (N) \Big] ^{-1} \qV (N, a_0) \: .
\eea
Here $\gV_0 (N)$ is related to the matrix of the
LO splitting function
$\PV _0(x) $ as in eq.~(\ref{zzz}), and as before $a_0 = a_s(Q^2_0)$.
The singlet evolution matrices $\UV_l (N)$ can be expressed in terms of
the anomalous dimensions $\gV_{k \leq l} (N)$. Technical details can be
found in ref.~\cite{BRVup}.

\begin{figure}[hbt]
\vspace{-14mm}
\begin{center}
\mbox{\hspace*{-2mm}\epsfig{file=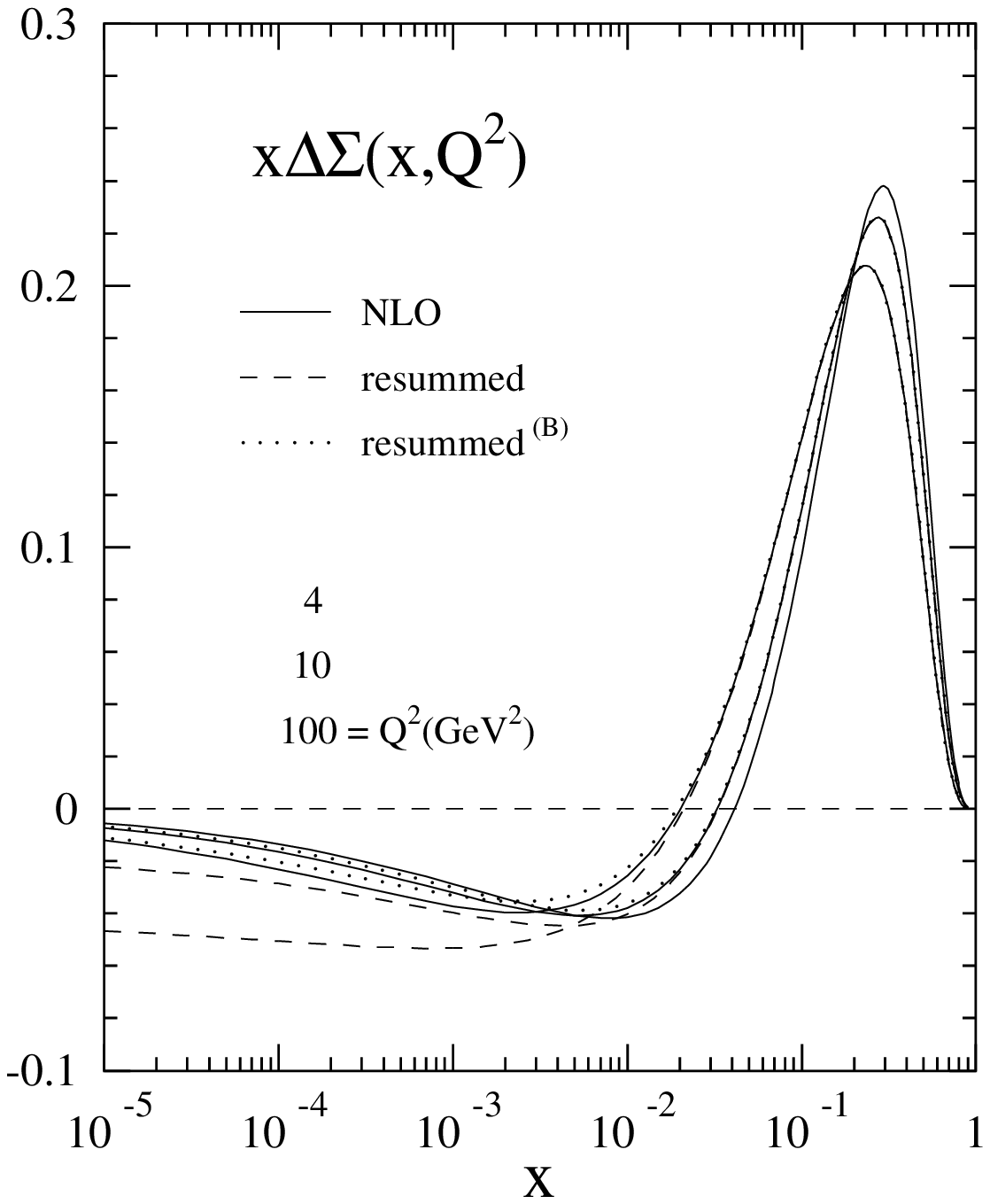,width=7.7cm}}
\vspace{-8mm}
\end{center}
{\sf {\bf Figure 7:}~~The evolution of the polarized singlet combination
$x\Delta \Sigma$ in NLO and including the resummed kernels.  The impact 
of possible subleading terms is illustrated by the prescription `(B)' 
in eq.~(\ref{xxx}). The input densities are from ref.~\cite{GRSV}.}
\vspace*{-1mm}
\end{figure}

The numerical consequences of the resummation in Section 3.2 on the
polarized parton densities and the structure functions $g_1^{\, ep}$ and
$g_1^{\, en}$ have been investigated in ref.~\cite{BVplb2}. In Figures 7
and 8 the results are displayed for the polarized singlet and gluon
densities, $x\Delta \Sigma $ and $x\Delta g $, respectively. The
initial distributions at $Q_0^2 = 4 \mbox{ GeV}^2$ have been adopted
from the GRSV `standard' parametrization \cite{GRSV}. The evolution
has been performed for $ \Lambda_{\overline {\rm MS}} (N_f = 4) = 200
\MeV$, and -- different from all other cases shown in this paper -- with
only three active quark flavours in the splitting functions
\cite{GRSV,GRV94}, i.e. $\Delta \Sigma $ is given by
\beq
 \Delta \Sigma =  \Delta u + \Delta \bar{u} + \Delta d + \Delta \bar{d}
		+ \Delta s + \Delta \bar{s} \: .
\eeq
As in the corresponding non--singlet `--'-case, the expanded solution
(\ref{Ssi}) represents an asymptotic series, which diverges
at very small values of $x$. For all the experimentally relevant
cases, $ x \gsim 10^{-5}$, retaining 8 -- 10 terms in eq.~(\ref{Ssi}) 
is however adequate for obtaining accurate results.

\begin{figure}[hbt]
\vspace*{-11mm}
\begin{center}
\mbox{\hspace*{-2mm}\epsfig{file=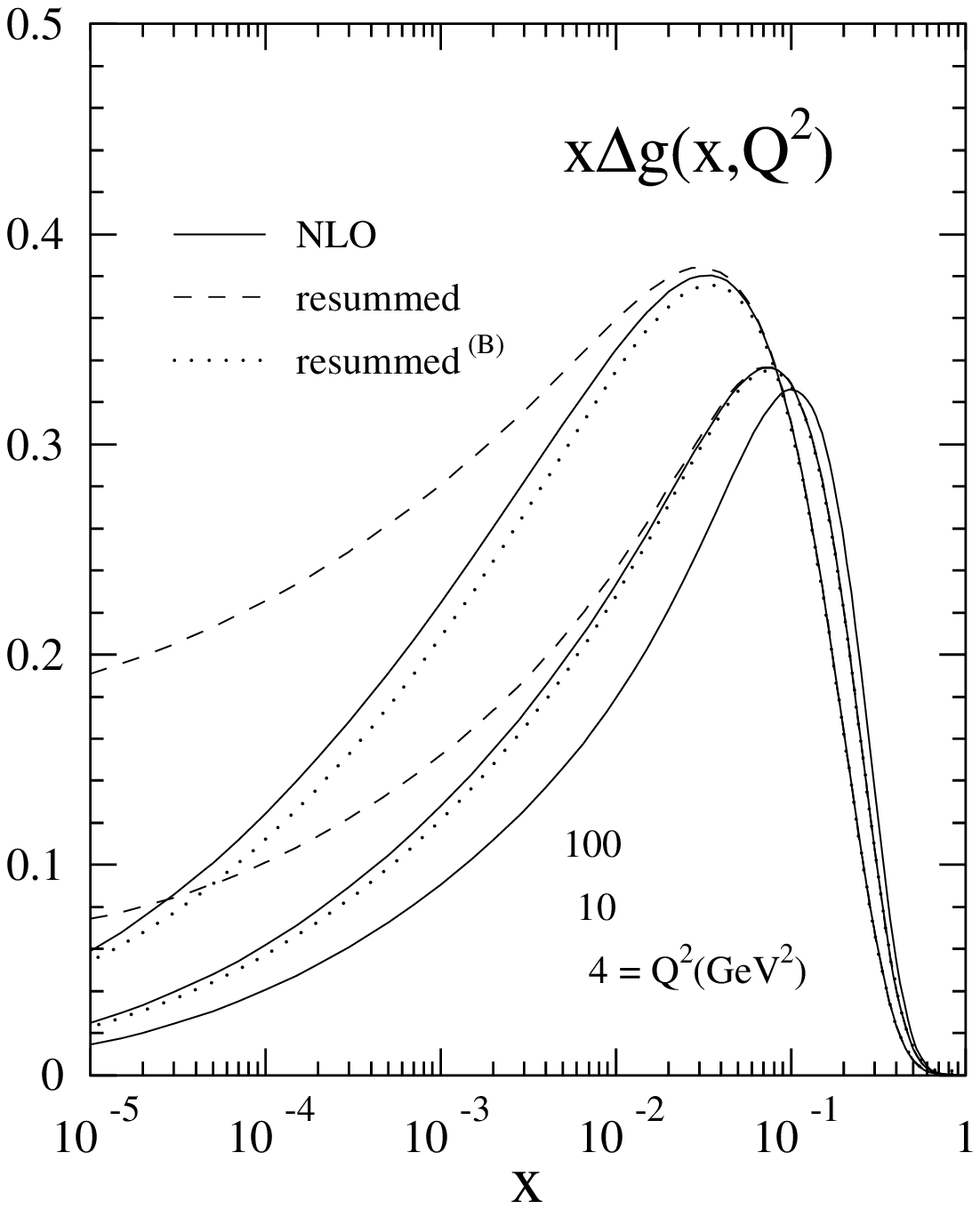,width=7.7cm}}
\vspace{-7mm}
\end{center}
{\sf {\bf Figure 8:}~~As in Figure 7, but for the polarized gluon
momentum distribution $x \Delta g $. As in the previous figure, the
$Q^2$-values in the legend are ordered according to the sequence of the
curves at small $x$.}
\vspace*{-3mm}
\end{figure}

Figures 7 and 8 show that the resummation effects are much larger for
$\Delta \Sigma $ and $\Delta g $ than for the non--singlet quantities
considered in Section 4.2, as to be expected from the comparison of the
expansion coefficients in Tables~1 and 2. E.g., the ratio of the
(unsubtracted) resummed results to the NLO evolution amounts to about
1.72 (1.64) for $\Delta \Sigma $ ($\Delta g $) at $Q^2 = 10 
\mbox{ GeV}^2$ and $x = 10^{-4}$.

Also illustrated in these figures [by the results for the prescription
`(B)' in eq.~(\ref{xxx})] is the possible impact of the yet uncalculated
terms in the higher--order anomalous dimensions which are down by one
power of $N$ with respect to the resummed leading pieces as $N \ra 0$.
The effect of these additional terms can be very large, in the present
example the resummation correction beyond NLO is practically cancelled.

\begin{table}[htb]
\vspace*{-10mm}
\begin{center}
\begin{tabular}{||c|c|c|c|c||}
\hline \hline
\multicolumn{1}{||c}{ } &
\multicolumn{2}{|c|}{ } &
\multicolumn{2}{c||}{ } \\[-0.3cm]
\multicolumn{1}{||c}{$Q^2$                       } &
\multicolumn{2}{|c|}{$      10  \mbox{ GeV}^2$   } &
\multicolumn{2}{c||}{$      100 \mbox{ GeV}^2$   } \\
\multicolumn{1}{||c}{ } &
\multicolumn{2}{|c|}{ } &
\multicolumn{2}{c||}{ } \\[-0.3cm]
\hline
 & & & & \\[-0.3cm]
\multicolumn{1}{||c|}{$x$} &
\multicolumn{1}{c|} {$10^{-4}$} &
\multicolumn{1}{c|} {$10^{-3}$} &
\multicolumn{1}{c|} {$10^{-4}$} &
\multicolumn{1}{c||}{$10^{-3}$} \\
 & & & & \\[-0.4cm]
\hline \hline
 & & & & \\[-0.3cm]
                 & -0.0100  & -0.0169  & -0.0171  & -0.0218  \\
$x \Delta\Sigma $& -0.0285  & -0.0396  & -0.0505  & -0.0523  \\
                 & -0.0473  & -0.0560  & -0.0855  & -0.0772  \\
\hline
 & & & & \\[-0.3cm]
                 &  0.019   &  0.034   &  0.053   &  0.071   \\
$x \Delta g $    &  0.101   &  0.152   &  0.226   &  0.281   \\
                 &  0.201   &  0.294   &  0.432   &  0.528   \\
\hline \hline
\end{tabular}

\end{center}
{\sf {\bf Table 3:}~~A comparison of the resummed evolution of the
polarized parton distributions for different assumptions on the gluon
distribution $\Delta g$. Upper lines: minimal gluon, middle lines:
standard set, lower lines: maximal gluon (and corresponding quark
distributions) of ref.~\protect\cite{GRSV} at $Q_0^2 = 4 \mbox{ GeV}^2
$.}
\vspace*{-6mm}
\end{table}

The small-$x$ evolution depends  strongly also on the virtually unknown
\cite{polrev} gluon input density. This is obvious from Table~3, where
the resummed results of Figures~7 and 8 are compared at two
representative values of $x$ and $Q^2$ to those obtained by evolving in 
the same way the `minimal $\Delta g $' and `maximal $\Delta g$' 
distributions of GRSV \cite{GRSV}. The variations are up to a factor of 
almost 5 (10) for $\Delta \Sigma $ ($\Delta g $), respectively. Thus 
both the unknown less singular terms in the anomalous dimensions and 
the present bounds on $\Delta g$, which are rather weak still, are the
dominant sources of uncertainty at small $x$.

\subsection{The unpolarized singlet case}
\label{sect44}

\vspace*{1mm}
\noindent
The solution of the evolution equations for the unpolarized singlet 
parton densities is analogous to the polarized case considered in the
previous section, see eq.~(\ref{Ssi}). The present case is special
-- and has therefore attracted much interest over the past years --
since only here precise measurements have been performed for small $x$, 
at HERA \cite{EXPrev}.
The quantitative impact of the resummation discussed in Section 3.3 has 
been studied for parton distributions and structure functions in 
refs.~\cite{EHW} and \cite{BRVup}. The latter analysis confirms and
extends the former one. Related investigations have been carried out in 
refs.~\cite{BF,FRT}.

Below the results are shown for initial distributions which, although
representing a somewhat simplified input, incorporate all features
relevant to this study in a sufficiently realistic way, especially the
small-$x$ powers as supported by HERA structure function data 
\cite{EXPrev}. Specifically, we take in the DIS factorization scheme
at $Q_0^2 = 4 $ GeV$^2$: 
\begin{eqnarray}
  xu_v \! &\! =\! & A_u x^{0.5} (1-x)^3, \:\: xd_v \, = \, A_d x^{0.5}, 
  (1-x)^4 \nonumber\\
  xS &\! =\! & \Sigma \, - \, xu_v \, - \, xd_v~~~~ = \, A_S x^{-0.2}, 
  (1-x)^7 \nonumber\\
  xg &\! =\! & A_g x^{-0.2} (1-x)^5, \:\: xc \, = \, x\bar{c} \: = \: 0 
  \: .  \nonumber\\
\end{eqnarray}
The evolution is performed for four active (massless) flavours, using 
$\Lambda_{\overline{\rm MS}}(N_f\! =\! 4) = 250$ MeV. The 
(SU(3)--symmetric) sea is assumed to carry 15\% of the proton's momentum
at the input scale; together with the sum rules this fixes the 
prefactors $A_i$.

\begin{figure}[htb]
\vspace{-2mm}
\begin{center}
\mbox{\hspace*{-3mm}\epsfig{file=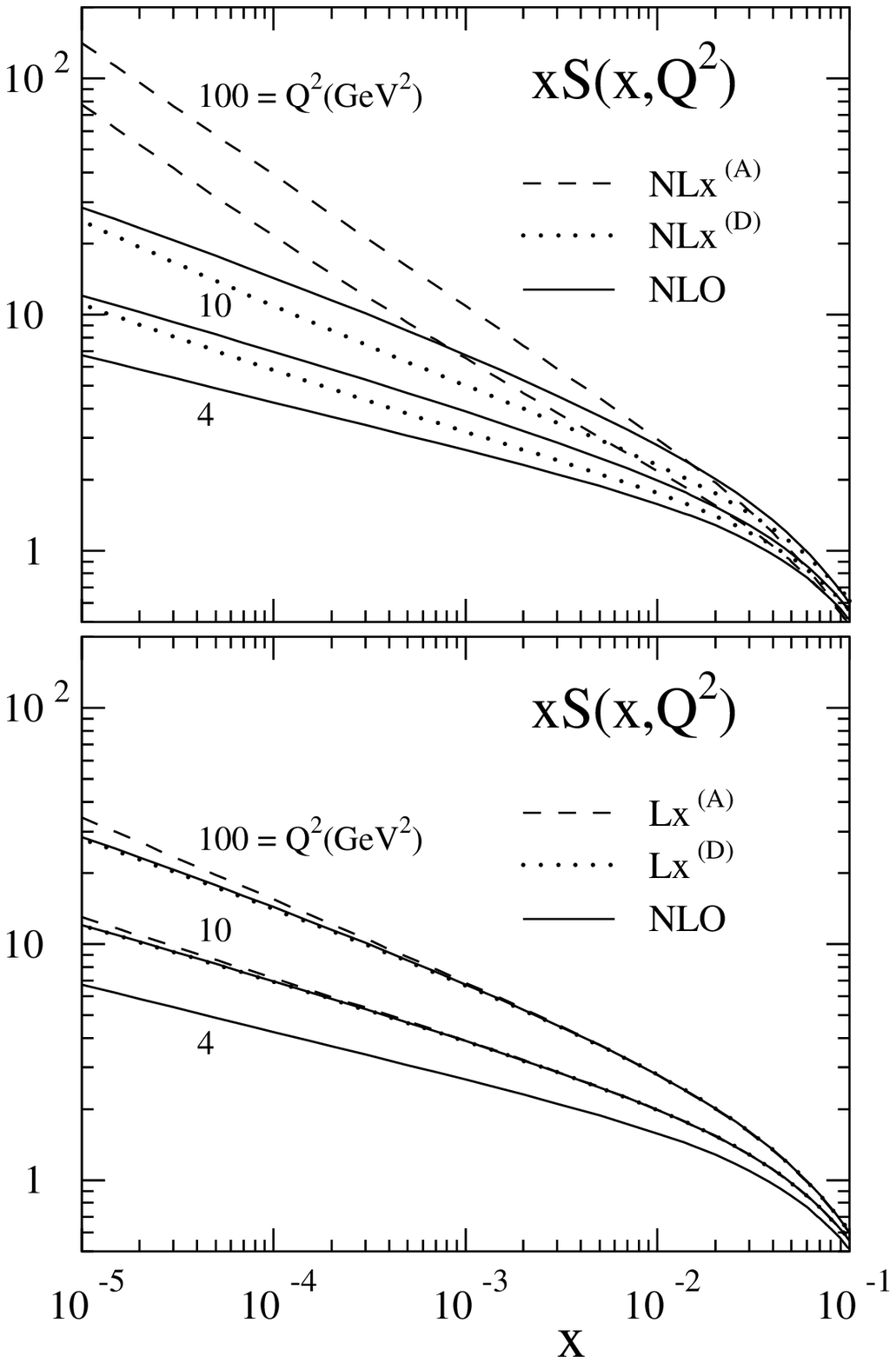,width=8.0cm}}
\vspace{-8mm}
\end{center}
{\sf {\bf Figure 9:}~~The small-$x$ evolution of the total unpolarized 
 sea quark density $xS$ including the resummed {\it Lx\/} \cite{LIPAT} 
  and {\it NLx\/} kernels \cite{CH} as compared to the NLO results. Two 
 prescriptions for implementing the energy--momentum sum rule have been 
 applied, c.f.\ Section 3.4.}
\vspace*{-8mm}
\end{figure}

\begin{figure}[htb]
\vspace*{-2mm}
\begin{center}
\mbox{\hspace*{-2mm}\epsfig{file=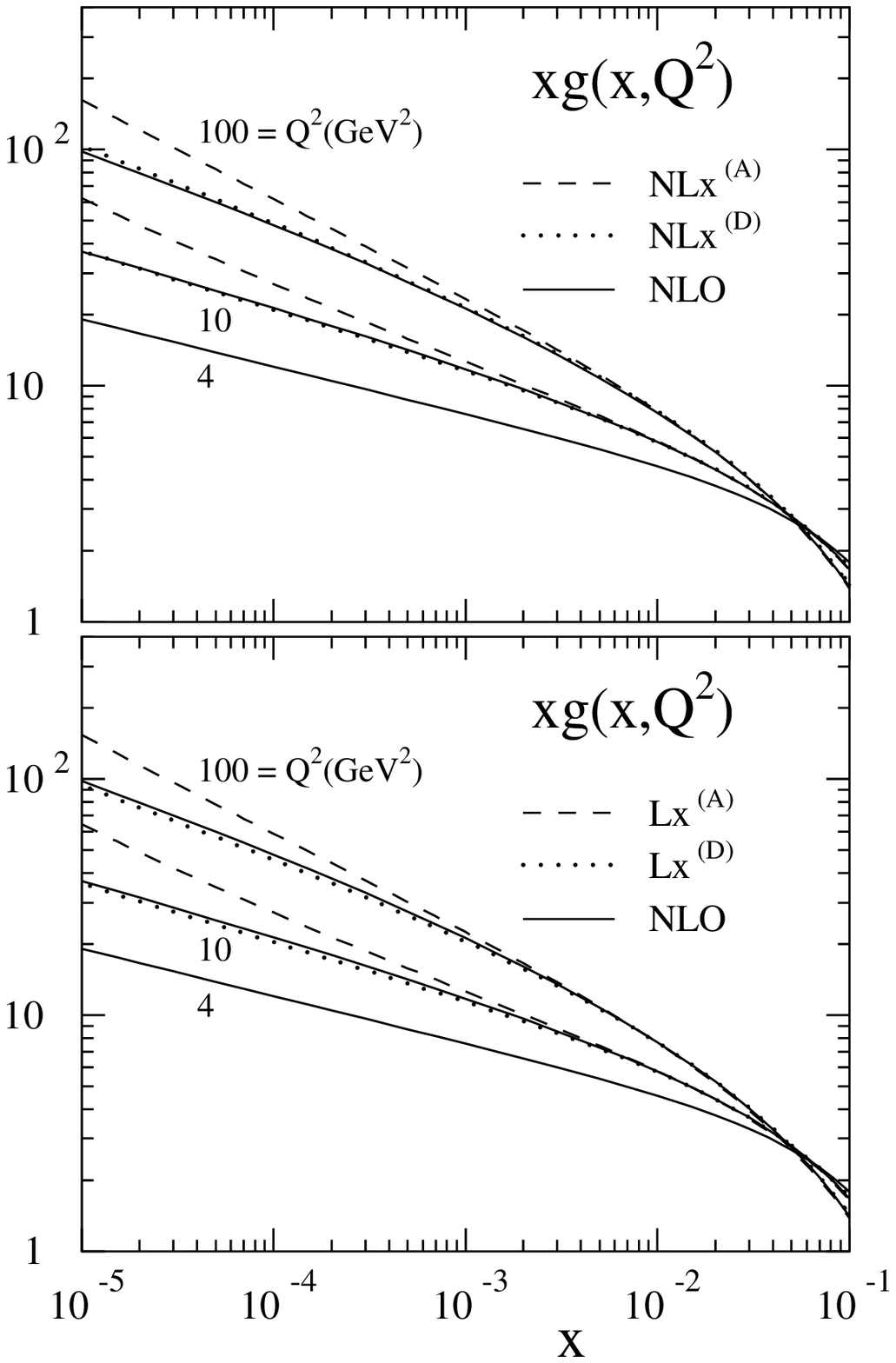,width=8.0cm}}
\vspace{-8mm}
\end{center}
{\sf {\bf Figure 10:}~~As in Figure 9, but for the unpolarized gluon 
 momentum distribution $xg$.}
\vspace*{-4mm}
\end{figure}

Figures 9 and 10 compare the resummation results separately for the {\it
 Lx\/} \cite{LIPAT} and {\it NLx\/} \cite{CH} series (c.f.\ Section 3.3) 
to the standard NLO evolution. In the {\it Lx\/} case, as expected from 
the matrix structure in eq.~(\ref{GAML}), the main effect is exerted on 
the gluon density $xg$. 
The impact on the quark evolution is rather moderate. Specifically, the
ratio to the NLO results amounts to about 1.3 (1.03) for $xg$ ($xS$), 
respectively, at $Q^2 = 10 $ GeV$^2$ and $x = 10^{-4}$, taking the
prescription `(A)' of eq.~(\ref{xxx}) for restoring the energy--momentum
sum rule. Including also the {\it NLx\/} quark terms \cite{CH}, on the 
other hand, results only in a small further modification of the gluon 
evolution, whereas the quark distributions are drastically affected. 
The ratios to the NLO results now read 1.3 (3.1) for $xg$ ($xS$) under 
the same conditions as before.  

A flavour of the possible importance of presently unknown less singular
terms in the higher--order anomalous dimensions is provided by the 
difference of the results of the choices `(A)' and `(D)' in 
eq.~(\ref{xxx}). Such terms can be vitally important, like in the 
polarized case studied in the previous section. As obvious from the 
figures not even the sign of the deviation from the NLO evolution can 
be taken for granted.

\vspace*{5mm}
\subsection{QED non--singlet radiative corrections}
\label{sect45}

\vspace*{1mm}
\noindent
The resummation of the $O(\alpha \ln^2 x)$ terms may also yield 
non--negligible contributions to QED radiative corrections 
\cite{BVcrac}. This has been investigated recently for the case of 
initial--state radiation in deep inelastic $eN$ scattering. In the 
range of large $y$ the effect can reach around 10\% of the differential 
Born cross section~\cite{BRVqed}, see Figure~11. These terms are not 
covered by the higher order resummations studied so far~\cite{KMSP,BRAD}
and reduce their effect~\cite{BRVqed}. Yet the complete NLO corrections 
for this process are not known and the size of less singular terms at
$O(\alpha^2)$ and their impact on the QED corrections is still to be 
determined (see, however, eq.~(\ref{KQED})).

For the corresponding corrections to $\sigma(e^+e^- \rightarrow 
\mu^+ \mu^-)$ near the $Z$-peak the situation is different. Since there 
the QED correction are completely known up to $O(\alpha^2)$~\cite{BBN},
a correction due to the resummation of the $O(\alpha^{l+1} \ln^{2l} x)$
terms is only necessary for the terms beyond two loop order, as in the
QCD cases considered before.  A numerical study of these effects can 
also be found in ref.~\cite{BRVqed}.
\begin{figure}
\vspace{-12mm}
\begin{center}
\mbox{\hspace*{-2mm}\epsfig{file=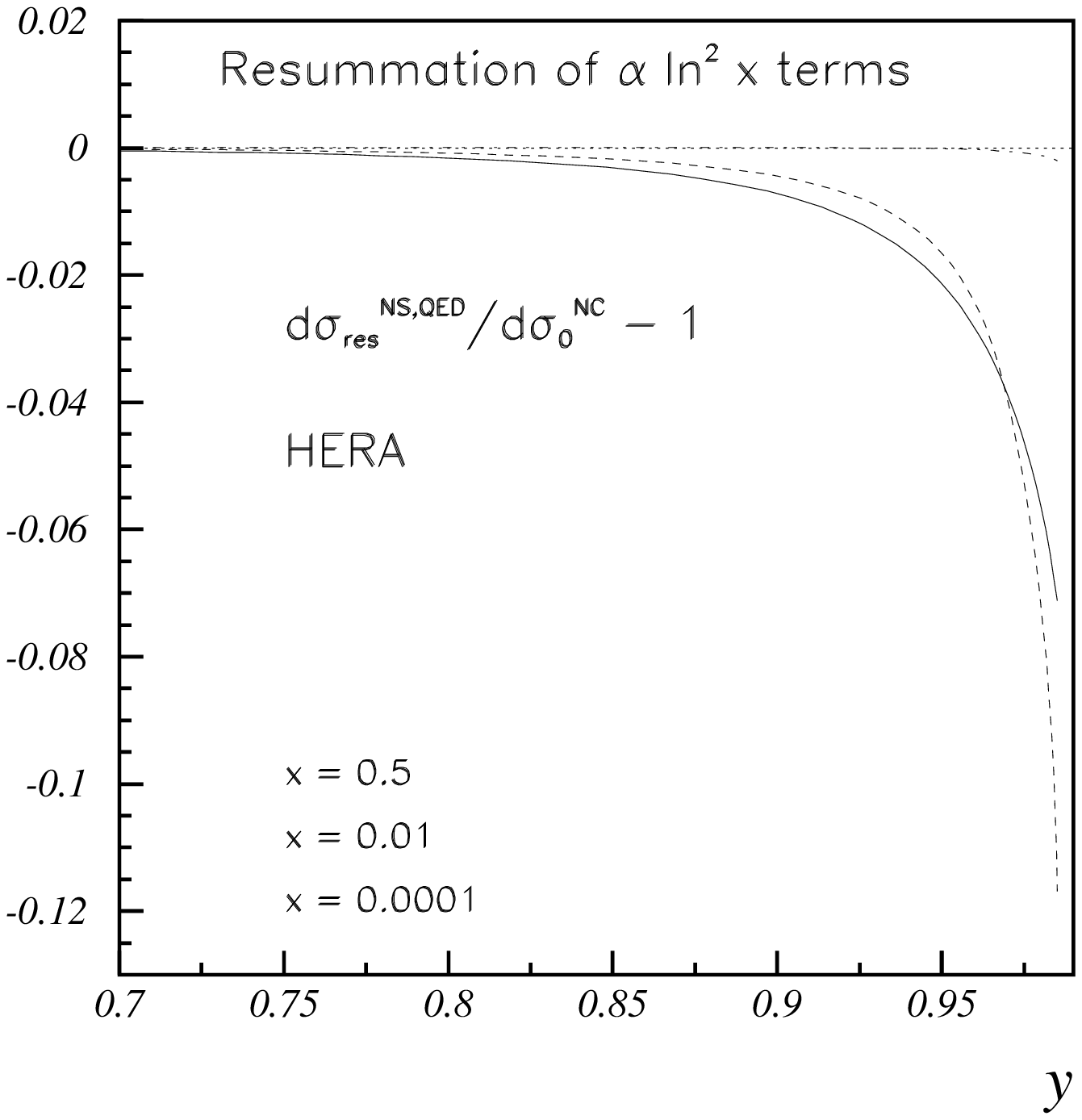,width=8.6cm}}
\vspace*{-9mm}
\end{center}
{\sf {\bf Figure 11:}~~The non--singlet (NS) resummed contribution of 
the $O(\alpha(\alpha \ln^2 x)^l)$ terms for the QED initial--state
correction to the neutral current deep--inelastic scattering cross 
section at HERA. The results are shown normalized to the differential
Born cross section.}
\vspace{-6mm}
\end{figure}

\section{Summary}

\noindent
The resummations of the leading small-$x$ terms in both unpolarized and
polarized, non--singlet and singlet anomalous dimensions have been
discussed. At NLO the results agree with those found for the most 
singular terms as $x \rightarrow 0$ in fixed--order calculations.
The so--called supersymmetric relation is satisfied by the results for 
the most singular small-$x$ terms to all orders, again for both the 
unpolarized and polarized cases.
These resummations allow the
prediction of
 the leading $x \rightarrow 0$
contributions to the 3--loop (NNLO) anomalous dimensions in the 
$\overline{\rm MS}$ scheme~\cite{CH,BVplb1,BVcrac,BVplb2}.
The coefficient functions are less singular for the non--singlet and
polarized singlet cases up to NNLO, $O(\alpha_s^2)$.

For the non--singlet structure functions the corrections due to the
$\alpha_s(\alpha_s \ln^2x)^{l}$ contributions are about $1\%$ or smaller
in the kinematical ranges probed so far and possibly accessible at HERA
including polarization~\cite{BVplb1,BVcrac}.
The non--singlet QED corrections in deep--inelastic scattering resumming
the $O(\alpha \ln^2 x)$ terms can reach values of up to 10\% at $x
\approx 10^{-4}$ and $y>0.9$~\cite{BRVqed}.

In the singlet case very large corrections are obtained for both
unpolarized and polarized parton densities and structure functions 
\cite{BVplb2,EHW,BRVup}.
As in the non--singlet cases possible less singular terms in the
higher order anomalous dimensions, however, are hardly suppressed 
against the presently resummed leading terms in the evolution: even a 
full compensation of the resummation effects cannot be excluded.

To draw firm conclusions on the small-$x$ evolution of singlet structure
functions also the next less singular terms have to be calculated. 
Since contributions even less singular than these ones may still cause 
relevant corrections, it appears to be indispensable to compare the 
corresponding results to those of future complete three--loop 
calculations.

\vspace{5mm}
\noindent
{\bf Acknowledgements :}
We would like to thank W. van Neerven and T. van Ritbergen for useful
discussions.  This work was supported in part by the EC Network 
`Human Capital and Mobility' under contract No.\ CHRX--CT923--0004, 
and by the German Federal Ministry for Research and Technology (BMBF) 
under contract \mbox{No.\ 05 7WZ91P (0).}

\end{document}